\documentstyle[preprint,aps]{revtex}
\renewcommand{\theequation}{\arabic{section}.\arabic{equation}}
\tighten
\begin{document}
\draft 
\title{Vector Positronium States in three-dimensional QED} 
\author{T.\ W.\ Allen\ and C.\ J.\ Burden \vspace*{0.2\baselineskip}}
\address{
Department of Theoretical Physics,
Research School of Physical Sciences and Engineering,
Australian National University, Canberra, ACT 0200, Australia
\vspace*{0.2\baselineskip}\\}
\maketitle
\begin{abstract}
The homogeneous Bethe-Salpeter equation is solved in the
quenched ladder approximation for the vector positronium states 
of 4-component quantum electrodynamics in 2 space 
and 1 time dimensions.  Fermion propagator input is from 
a Rainbow approximation Dyson-Schwinger solution, with a broad range of 
fermion masses considered.  This work is an extension of earlier work 
on the scalar spectrum of the same model.  The non-relativistic limit 
is also considered via the large fermion mass limit.  Classification 
of states via their transformation properties under discrete parity 
transformations allows analogies to be drawn with the meson spectrum 
of QCD.  
\end{abstract}
%-----------------------------------------------------------------
\section{Introduction}
This paper extends our previous 
nonperturbative studies of Quantum Electrodynamics in three spacetime 
dimensions (QED$_3$) ~\cite{Bu92,AB96} from scalar to vector positronium 
states.  

Being a confining theory, the low 
energy behaviour of QED$_3$ must be dealt with nonperturbatively.  
We consider QED$_3$ to be a simple but effective testing ground for 
non-perturbative methods commonly applied to Quantum Chromodynamics 
(QCD$_4$).  Again our approach is via a solution to the 
homogeneous Bethe-Salpeter equation (BSE) with fermion propagator
input from the Dyson-Schwinger equation (DSE).  We consider a 
numerically solvable system of integral equations within the quenched, 
ladder approximation for the BSE and the quenched, rainbow approximation 
for the DSE.  It is well known that the combination of ladder BSE and 
rainbow DSE respects Goldstone's theorem, i.e, in the chiral limit of 
zero current fermion mass, the spectrum admits a massless 
``pion''~\cite{DS79}.  Our four-component fermion version of QED$_3$ admits a 
spontaneously broken chiral-like symmetry leading to a doublet of 
Goldstone bosons.  It is reasonable to expect therefore that the 
important effect of chiral symmetry breaking on the light bound state 
spectrum will be well modelled by this approximation.  

We note, however, that our truncation does break local gauge covariance.  
Of particular concern is the effect of the rainbow approximation on the 
analytic structure of the fermion propagator.  This approximation is known 
to generate ghost poles in the complex momentum plane in both 
QED$_4$~\cite{M94} and QED$_3$~\cite{M95}.  Such poles can be an impediment
to the BSE formulated in Euclidean space, which samples the fermion
propagator over a region of the complex momentum plane~\cite{SC92}.  
Herein we assume the position that, provided the ghost poles do not
impinge upon this region of the complex plane, the rainbow-ladder
approximation is a reasonable one.  We acknowledge, however, that this
point deserves further investigation.  Several QED$_3$ studies~\cite{Sev}
have examined the effect on the spacelike behaviour of the fermion propagator
if the fermion-photon vertex is replaced by a more sophisticated ansatz
satisfying the Ward-Takahashi identity.  These studies find the spacelike
behaviour to be qualitatively similar to that of the rainbow approximation.
Extension of these results to the remainder of the complex plane is
beyond the scope of this current paper.  We also note that any improvement
in the fermion-photon vertex should be matched by a corresponding 
improvement in the Bethe-Salpeter (BS) kernel in such a way as to preserve
Goldstone's theorem.  That this can in principle be achieved by
consistently matching a loop expansion of the vertex with crossed diagrams
in the BS kernel has recently been demonstrated~\cite{BRS96}.

The extension from scalar to vector states enriches considerably the 
spectrum of bound states.  This facilitates an understanding of the 
importance of symmetry principles in determining the dynamics of the 
positronium spectrum, and allows analogies to be drawn with 
the meson spectrum of QCD$_4$.  We find that there is 
a one to one correspondence between the classification of states in 
4-component QED$_3$ in terms of ``axial parity'' and charge conjugation on 
the one hand, and the conventional $J^{PC}$ classification of charge neutral 
mesons in QCD$_4$ on the other.  

We consider bare fermion mass ranging from zero to large values.  For 
large fermion masses we are able to make contact with 
the non-relativistic limit.  In this limit we extend our previous derivation 
of the Schr\"odinger equation as a limit of the BSE formalism to higher spin 
states, and observe spectrum degeneracies analogous to those of heavy quark 
effective theory.  

The paper is organised as follows.  
In section II we look at the Bethe-Salpeter and Dyson-Schwinger 
approximations used and set out the method we employ to find the
vector bound state masses.  
Transformation properties in QED$_3$, with special attention 
given to the newly considered vector states, are discussed in appendix A. 
These transformation properties are necessary for 
understanding the structure of Bethe-Salpeter amplitudes, 
and classification of the vector bound states.
The Bethe-Salpeter coupled integral equations for the vector states are 
given in appendix B.  
Section III describes the nonrelativistic limit for the vector states.
In section IV numerical vector Bethe-Salpeter
solutions are reported and comparisons are made with existing
nonrelativistic limit calculations.  The results are discussed 
and conclusions drawn in section V.

%------------------------------------------------------------------
\setcounter{equation}{0}
\section{Solving the Bethe-Salpeter Equation for Vector States}

As in previous work~\cite{AB96}, the BS kernel 
is the quenched ladder approximation kernel (bare one-photon exchange).
Again, for convenience, we use Feynman gauge and work with the Euclidean 
metric.  The BSE can then be written as
\begin{equation}
\Gamma_{\nu}(p,P) = -e^2 \int \mbox{$ \, \frac{d^3q}{(2\pi)^3} \,$} 
D(p-q) \gamma_\mu S(\mbox{$ \, \frac{1}{2} \,$} P+q)
       \Gamma_{\nu}(q,P)S(-\mbox{$ \, \frac{1}{2} \,$} P+q) 
       \gamma_\mu, \label{eq:BS}
\end{equation}
where $\Gamma_{\nu}(p,P)$ is the one fermion irreducible 
positronium-fermion-antifermion vertex with external legs 
amputated.  The photon 
propagator $D(p-q)$ in Feynman gauge is $1/(p-q)^2$.  The
fermion propagator $S$ is the solution to a truncated DSE.  
For bare fermion mass $m$ this truncated DSE (the quenched Rainbow 
Approximation) is
\begin{equation}
\Sigma(p)= S(p)^{-1} - (i\!\not \! p + m) = 
 e^2 \int \mbox{$ \, \frac{d^3q}{(2\pi)^3} \,$} D(p-q) \gamma_\mu 
 S(q) \gamma_\mu.  \label{eq:DS}
\end{equation}
From here on we use the units $e^2 = 1$ allowable due to the fact that in
the quenched approximation the BSE and DSE can be recast in terms of
dimensionless momentum $p/e^2$ and mass $m/e^2$.

We write the fermion propagator in the following general form
\begin{equation}
S(p)= -i \not\! p \sigma_V(p^2)+\sigma_S(p^2)  \;\;\;\;\;  \mbox{or} \;\;\;\;\;  
S(p)= \frac{1}{ i \not\! p A(p^2)+B(p^2) },
\label{eq:DSEF}
\end{equation}
where the vector and scalar parts of the propagator are given by
\begin{equation}
\sigma_V(p^2)= \frac{A(p^2)}{p^2 A^2(p^2) + B^2(p^2)},  \;\;\;\;\;  \mbox{and} \;\;\;\;
\sigma_S(p^2)= \frac{B(p^2)}{p^2 A^2(p^2) + B^2(p^2)}.
\label{eq:SIGDEF}
\end{equation}
Non-zero $B$ signals dynamical fermion mass generation in the
massless limit.  Suitable analytic fits to $A$ and $B$ were found in earlier
work~\cite{AB96}, namely 
$$
A_{\rm fit}(p^2)= \frac{a_1}{(a_2^{\,2} +p^2)^{\frac{1}{2}}}
             +a_3 e^{-a_4p^2} +1,
$$
\begin{equation}
B_{\rm fit}(p^2)= \frac{b_1}{b_2 +p^2}+b_3 e^{-b_4p^2} + m.  \label{eq:FIT}
\end{equation}
The parameters $a_n$,$b_n$ (functions of fermion mass) were the result
of fits to iterative solutions to Eq.~(\ref{eq:DS}) and can be
found in the aforementioned paper.  
The analytic properties of the propagator are summarised at the end of 
this section.

In order to solve the BSE we must write it as a set
of numerically tractable coupled integral equations.  
To do this, we write the bound state amplitude $\Gamma_{\nu}$ in its most 
general form consistent with
the parity and charge conjugation of the required bound
state, substitute it into the BSE and project out the coefficient 
functions for the individual Dirac components. 
The general vector vertex can be found in appendix A and the integral
equations in appendix B.

The solution to the BSE involves iteration of the eight coupled integral 
equations Eq.~(\ref{eq:IE}).  These equations can be written in
the form of an eigenvalue problem.  
Let ${\bf f}=(a,b,c,d,e,f,g,h)^{\rm T}$ then we have  
\begin{equation}
\int dq_3
 \int d\mbox{$\left| {\bf q} \right|$} \, K(\mbox{$\left| {\bf p} 
 \right|$},p_3;\mbox{$\left| {\bf q} \right|$},q_3;M) {\bf f}
 (\mbox{$\left| {\bf q} \right|$},q_3;M)
 =  \Lambda(M) \,\,\, {\bf f}(\mbox{$\left| {\bf p} \right|$},p_3;M), \label{eq:EV}
\end{equation}
for a given test mass $M$.  This equation is solved for different test 
bound state masses until an eigenvalue $\Lambda(M)=1$ is obtained.  Such
an equation exists for each symmetry case 
(vector ${\cal C}= -1$, vector ${\cal C}= +1$, axivector 
${\cal C}=-1$ and axivector ${\cal C}=+1$) and for each fermion mass $m$. 

The BSE described in this section requires a fermion propagator input in the 
form of Eq.~(\ref{eq:DSEF}) and this needs to be available over a region 
in the complex momentum plane defined by $Q^2$ from Eq.~(\ref{eq:QDEF}) 
with $q_3$ and $\mbox{$\left| {\bf q} \right|$}$ real.  
This is the region~\cite{SC92}
\begin{equation}
\Omega = \left\{ Q^2 = X + iY \left| X > \frac{Y^2}{M^2} 
- \frac{1}{4} M^2 \right. \right\}.
\label{eq:REGION}
\end{equation}
The DSE solution should be well behaved over $\Omega$.  In previous 
work~\cite{AB96} we have studied the analytical properties of the propagator 
and here we summarise briefly.

Conjugate poles exist where the factor $p^2 A^2 + B^2$ 
appearing in the denominator of the fermion propagator is zero.  
Ref.~\cite{AB96} lists the conjugate poles arising from the fits for each 
fermion mass and the corresponding maximum bound state masses allowed.  The
maximum $M$ allowed is the value for which the boundary of 
$\Omega$ in Eq.~(\ref{eq:REGION}) coincides with the
conjugate poles as we should not allow the singularities to enter into
the BSE sampling region. If the singularities were to enter that region
it would be necessary for compensating zeros to exist in the Dirac
coefficients (ie: nodes in the wavefunction).
Note that the region boundary~(\ref{eq:REGION}) is bound state mass $M$
dependent and thus one must take care as $M$ increases during the bound state 
search.

We have shown~\cite{AB96} that both the DSE solution and our spacelike 
fits Eq.~(\ref{eq:FIT}) have conjugate singularities.  For small bare 
fermion mass $m$ we find that the fits reproduce well the position of 
singularities.  However, as $m$ is increased beyond the scale $e^2 = 1$ 
of the model, the spacelike propagator functions $A$ and $B$ become quite 
level, and the propagator poles recede further into the timelike half of 
the complex plane.  In this limit, numerical propagator fits are unable 
to model accurately the important analytic pole structure in the 
timelike half of the $p^2$-plane.  We therefore find our numerical bound 
state mass solutions contaminated by noise for $m>>1$.  For this reason, 
the non-relativistic limit $m\rightarrow \infty$ of the theory must be 
treated separately.  

%------------------------------------------------------------------
\setcounter{equation}{0}
\section{Nonrelativistic Limit}
In previous work~\cite{AB96} a nonrelativistic limit ($m\rightarrow \infty$)    
of the BSE was derived for the scalar states of the positronium system.
In this work we follow the same procedure for the vector states
which will allow a comparison to be made with the large fermion mass
limit of the full BSE calculation (solution to Eq.~(\ref{eq:IE})).  

Consider first the BSE, in which we set the bound state momentum
in Eq.~(\ref{eq:BS}) equal to
$P_{\mu}=(2m+\delta)iw_{\mu}$, where $w_{\mu}=(0,0,1)$ and
$-\delta$ is a ``binding energy''.  This gives (according to the
momentum distribution in Eq.~(\ref{eq:BS}))
\begin{equation}
\Gamma_{\nu}(p) = - \int \frac{d^3q}{(2\pi)^3} D(p-q) \, \gamma_\mu \, 
S\left[-\left(m+\frac{\delta}{2}\right)iw+q\right]
       \Gamma_{\nu}(q) \, S\left[\left(m+\frac{\delta}{2}\right)iw+q\right] \gamma_\mu. 
\label{eq:NRBSE}
\end{equation}
An expansion in orders of $1/m$~\cite{AB96} leads to the propagators
\begin{equation}
S\left[\left(m+\frac{\delta}{2}\right)iw+q \right]
 = \frac{(1+\gamma_3)/2}{ [ -(1/2) \delta+iq_3+ 
 \left|{\bf q}\right|^2 / 2m ] + \Sigma_{+}(q_3,\left|{\bf q}\right|)} \, 
 + \, O\left(\frac{\ln m}{m}\right)
\label{eq:SPLUS}
\end{equation}
and 
\begin{equation}
S\left[-\left(m+\frac{\delta}{2}\right)iw+q \right]
 = \frac{(1-\gamma_3)/2}{ [ -(1/2) \delta-iq_3+
 \left|{\bf q}\right|^2 / 2m ] 
 + \Sigma_{-}(q_3,\left|{\bf q}\right|) } \, 
 + \, O\left(\frac{\ln m}{m}\right).
\label{eq:SMINUS} 
\end{equation}
where, to one loop order in the self energy,
\begin{equation}
\Sigma_{\pm}(q_3,\left|{\bf q}\right|)
 = - \frac{1}{4\pi} \ln \left[ \frac{1}{2m} \left( -\frac{1}{2}\delta \pm iq_3
 + \frac{\left|{\bf q}\right|^2}{2m} \right) \right] \, 
 + \, O\left(\frac{1}{m^2}\right).
\label{eq:SG1LP} 
\end{equation}
In Ref.~\cite{AB96} we assumed that, if the fermion self energy is calculated
to all orders in rainbow approximation, the fermion self energy feeds back
into the loop integral via the propagator to replace Eq.~(\ref{eq:SG1LP})
by
\begin{equation}
\Sigma_{\pm}(q_3,\left|{\bf q}\right|)
 = - \frac{1}{4\pi} \ln \left[ \frac{1}{2m} \left( -\frac{1}{2}\delta \pm iq_3
 + \frac{\left|{\bf q}\right|^2}{2m} + \Sigma_{\pm}(q_3,\left|{\bf q}\right|) \right) 
 \right].
\label{eq:SGRAIN} 
\end{equation}
As before we assume that this equation provides an approximation to the 
rainbow DSE in the nonrelativistic limit. 

Since the vertex $\Gamma_{\nu}$ is defined with the fermion legs 
truncated, and $S \propto \frac{1}{2}(1 \pm \gamma_3)$, we find that the only
relevant part of $\Gamma_{\nu}$ is the projection 
$\mbox{$ \, \frac{1}{2} \,$}(1-\gamma_3) \, \Gamma_{\nu} \, 
\mbox{$ \, \frac{1}{2} \,$}(1+\gamma_3)$  
and with this in mind, the general vector form in Eq.~(\ref{eq:VERTEX}) 
becomes
\begin{eqnarray}
\mbox{$ \, \frac{1}{2} \,$}(1-\gamma_3) \, \Gamma_{\nu}^V(q,P) \, 
\mbox{$ \, \frac{1}{2} \,$}(1+\gamma_3) & =  &
\mbox{$ \, \frac{1}{2} \,$}(1-\gamma_3) \left[ \,\,\, 
 H_1(q_3,\mbox{$\left| {\bf q} \right|$}) \,\,  u_{\nu}(q)\not \! u(q)  
\right. \nonumber \\ &   &  \,\,\, \left.  
\,\,\,\,\,\,\,\,\,\,\,\,\,\,\,\,\,\,\,\,\,\,\, +
 H_2(q_3,\mbox{$\left| {\bf q} \right|$}) \,\,  v_{\nu}(q)\not \! v(q) 
\,\,\, \right]  \label{eq:VTMP}
\end{eqnarray}
where we have introduced the linear combinations $H_1=f_3+f_4/M$ and 
$H_2=f_8 + f_7 M$.
In the axivector case the general form in Eq.~(\ref{eq:AVS}) becomes
\begin{eqnarray}
\mbox{$ \, \frac{1}{2} \,$}(1-\gamma_3) \, \Gamma_{\nu}^{AV} \, 
\mbox{$ \, \frac{1}{2} \,$}(1+\gamma_3)  & = & 
\mbox{$ \, \frac{1}{2} \,$}(1-\gamma_3)  
\left(\begin{array}{c} \gamma_4 \\ \gamma_5\end{array} \right)  \left[ \,\,\,
 H_1(q_3,\mbox{$\left| {\bf q} \right|$}) \,\, u_{\nu}(q)  
 \right. \nonumber \\ &   &  \,\,\, \left.  \,\,\,\,\,\,\,\,\,\,
 \,\,\,\,\,\,\,\,\,\,\,\,\,\,\,\,\,\,\,\,\,\,\,\,\,\,\,\,\,\,\,\,\,\,
  +  H_2(q_3,\mbox{$\left| {\bf q} \right|$}) \,\, v_{\nu}(q) \gamma_{45} 
\,\,\, \right]
\label{eq:AVTMP}
\end{eqnarray}
where $H_1=f_1+if_2M$ and $H_2=f_5+if_6M$.

Substituting Eqs.~(\ref{eq:SPLUS}), (\ref{eq:SMINUS}) and
(\ref{eq:VTMP}) into Eq.~(\ref{eq:NRBSE}) one obtains for 
the vector states two integral equations (with 
$H_1^{'}=H_1/\mbox{$\left| {\bf q} \right|$}^2$ and 
$H_2^{'}=-H_2/M^2\mbox{$\left| {\bf q} \right|$}^2$)
$$
H_1^{'}(p)  =  \int \frac{d^3q}{(2\pi)^3} \, \frac{1}{(p - q)^2} 
\frac{\left[ H_1^{'}(q) \cos^2\theta+H_2^{'}(q)\sin^2\theta \right]}
{ \left|-\frac{1}{2}\delta+iq_3 + \frac{\left|{\bf q}\right|^2}{2m} 
+ \Sigma_{\pm}(q_3,\left|{\bf q}\right|)
 \right|^2 },  
$$
\begin{equation}
H_2^{'}(p)  =  \int \frac{d^3q}{(2\pi)^3} \, \frac{1}{(p - q)^2} 
\frac{\left[ H_1^{'}(q) \sin^2\theta+H_2^{'}(q)\cos^2\theta \right]}
{ \left|-\frac{1}{2}\delta+iq_3 + \frac{\left|{\bf q}\right|^2}{2m} 
+ \Sigma_{\pm}(q_3,\left|{\bf q}\right|) \right|^2 }.  
\label{eq:NRVTMP1}
\end{equation}
In these equations $\theta$ is the angle separating the 2-vectors
$\bf{p}$ and $\bf{q}$, and 
$H_{1,2}^{'}(q)=H_{1,2}^{'}(q_3,\mbox{$\left| {\bf q} \right|$})$.
It is clear that two independent equations
can be found by applying simple linear combinations.  Let $F=H_1^{'}+H_2^{'}$
and $G=H_1^{'}-H_2^{'}$, then 
\begin{equation}
F(p) =\int \frac{d^3q}{(2\pi)^3} \, \frac{1}{(p - q)^2} \frac{F(q)}
{ \left|-\frac{1}{2}\delta+iq_3 + \frac{\left|{\bf q}\right|^2}{2m} 
+ \Sigma_{\pm}(q_3,\left|{\bf q}\right|) \right|^2 }.
\label{eq:NRV1}
\end{equation}
This is identical to the non-relativistic limit of the 
axiscalar equation which has been solved numerically in 
Ref.~\cite{AB96}, and where only a ${\cal C}=+1$ solution could be
found (where $F$ is even in $q.P$).  
As we have seen, the function $F$ in the vector case involves a linear
combination of functions $f_3, f_4, f_7$, and $f_8$.  In appendix A we  
see that these are all odd in $q.P$ for the vector ${\cal C}=+1$ case
and even for the vector ${\cal C}=-1$ case.  Thus it is clear that the
axiscalar ${\cal C}=+1$ solution is degenerate, in the nonrelativistic
limit, with a vector ${\cal C}=-1$ state.

We also obtain a new equation, namely 
\begin{equation}
G(p)  =  \int \frac{d^3q}{(2\pi)^3} \, \frac{1}{(p - q)^2} \frac{\left[ 
2\left(\frac{\bf{p.q}}{\left|{\bf p}\right|\left|{\bf q}\right|} \right)^2 
-1 \right] \,\,  G(q)}
{ \left|-\frac{1}{2}\delta+iq_3 + \frac{\left|{\bf q}\right|^2}{2m} 
+ \Sigma_{\pm}(q_3,\left|{\bf q}\right|) \right|^2 } \, .
\label{eq:NRV2}
\end{equation}
Note that the constituents of $G$ used in the 
linear combination, namely $f_3$, $f_4$, $f_7$ and $f_8$, 
are all odd in $q.P$ for the ${\cal C} = +1$ case, and even in 
$q.P$ for the ${\cal C} = -1$ case, and so is the function $G$.  

In the axivector case we find that the surviving terms in
$\mbox{$ \, \frac{1}{2} \,$}(1-\gamma_3) \, \Gamma^{AV}_{\nu} \, 
\mbox{$ \, \frac{1}{2} \,$}(1+\gamma_3)$ can be collected 
into two linear combinations ($H_1(p)$ and $H_2(p)$ defined below 
Eq.~(\ref{eq:AVTMP})) 
and rescaled by dividing each by $\left|{\bf p}\right|$ to find two 
identical decoupled  
equations which are also identical to the scalar equation of previous work.  
If $F(p)=H_1(p)/\left|{\bf p}\right|$ or 
$F(p)=H_2(p)/\left|{\bf p}\right|$, we have
\begin{equation}
F(p) = \int \frac{d^3q}{(2\pi)^3} \,
     \frac{1}{(p - q)^2} 
     \frac{\bf{p.q}}{\left|{\bf p}\right|\left|{\bf q}\right|}
     \frac{F(q)}{ \left|-\frac{1}{2}\delta+
     iq_3 + \frac{\left|{\bf q}\right|^2}{2m} 
     + \Sigma_{\pm}(q_3,\left|{\bf q}\right|)
                        \right|^2 }
\label{eq:NRS}
\end{equation}
In one case $F$ is an even function of $q.P$ and in the other it is an 
odd function.   
The same equation~(\ref{eq:NRS}) could only be solved
with $F$ an even function of $q.P$ in previous work~\cite{AB96} where the 
scalar ${\cal C}=+1$ solution was found.  In the current case, solutions to 
Eq.~(\ref{eq:NRS}) can be identified with the axivector ${\cal C} = +1$ state 
by making the identification $F \propto H_1$, or with the axivector 
${\cal C} = -1$ state by making the identification $F \propto H_2$.  
We conclude that the axivector ${\cal C} = \pm 1$ states are mass 
degenerate in the $m \rightarrow \infty$ limit.  

We should therefore see, in the large fermion mass limit, a vector
state with ${\cal C}=-1$ degenerate with the axiscalar state with
${\cal C}=+1$, new vector states (possibly ${\cal C}=-1$ and ${\cal C}=+1$) 
which are solutions 
to Eq.~(\ref{eq:NRV2}), and axivector states with ${\cal C}=\pm 1$ which are 
degenerate with the scalar ${\cal C}=+1$ state.  The degeneracies 
between scalar and vector states are analogous to those of heavy quark 
effective theory \cite{N94}, in which the hadron is insensitive to the 
heavy quark spin to leading order in the inverse quark mass.  

Eqs.~(\ref{eq:NRV1}), (\ref{eq:NRV2}) and (\ref{eq:NRS}) correspond to 
states with orbital angular momentum $\ell = 0$, 2 and 1 respectively.  
To prove this, we show that these integral equations are equivalent to 
the Schr\"odinger equation of Koures in Ref.~\cite{Ko96}.  
Following the working of Ref.~\cite{AB96}, we see that Eqs.~(\ref{eq:NRV1}),
(\ref{eq:NRV2}) and (\ref{eq:NRS}) may be rewritten as in Eq.~(3.27) 
of Ref.~\cite{AB96}, namely 
\begin{equation} 
\left\{-\delta + \frac{\left|{\bf p}\right|^2}{m} + 
2\mbox{Re}\Sigma_{-}(p_3,\left|{\bf p}\right|)\right\}
 \Phi(p_3,\left|{\bf p}\right|) = \int \frac{d^2 {\bf q}}{(2\pi)^2} 
 V(\left|{\bf p - q}\right|)
 \Phi(q_3,\left|{\bf q}\right|) \chi({\bf p},{\bf q}) \label{eq:OUREQS}
\end{equation}
where $\chi({\bf p},{\bf q})$ is one of $1$, 
$\frac{{\bf p.q}}{\left|{\bf p}\right| \left|{\bf q}\right|}$ or
$\left\{ 2\left(\frac{{\bf p.q} }
   { \left|{\bf p}\right| \left|{\bf q}\right|} \right)^2 - 1 \right\}$.
In the final analysis the self energy $\Sigma_{-}$ serves the purpose of 
cancelling infrared divergence arising from the logarithmic potential $V$.  

On the other hand, consider the Schr\"odinger equation for a particle of 
orbital angular momentum $\ell$ in $(2 + 1)$ dimensions \cite{Ko96}: 
\begin{equation} 
\left\{-\delta + \frac{\nabla^2}{m} + 2\mbox{Re}\Sigma_{-}\right\}
\left[ \tilde{\phi}(\left|{\bf r}\right|) e^{\pm i \ell \theta} \right] =  
V(r) \left[ \tilde{\phi}(\left|{\bf r}\right|) 
e^{\pm i \ell \theta} \right]. \label{eq:SCHROL1}
\end{equation}
One easily shows that 
\begin{equation} 
\mbox{F.T. of } 
\left[ \tilde{\phi}(\left|{\bf r}\right|) e^{\pm i \ell \theta} \right] =  
\int d^2 {\bf r} \tilde{\phi}(\left|{\bf r}\right|) 
         e^{i{\bf r}.{\bf p} \pm i\ell\theta} = 
  \phi(\left|{\bf p}\right|) e^{\pm i \ell \theta_p}, 
\label{eq:FT1}
\end{equation}
where we have defined ($J_\ell$ is Bessel function of order $\ell$)
\begin{equation} 
\phi(\left|{\bf p}\right|) = 2 \pi i^{\ell} \int_{0}^{\infty} 
J_\ell(\left|{\bf p}\right| \left|{\bf r}\right|) 
\tilde{\phi}(\left|{\bf r}\right|)
\left|{\bf r}\right| d\left|{\bf r}\right|, 
\label{eq:PHIP}
\end{equation}
and $\theta$ is the angle the vector ${\bf r}$ makes with the $r_1$-axis, 
$\theta_p$ is the angle the vector ${\bf p}$ makes with the $p_1$-axis.  
From this we find that the F.T of the $\mbox{r.h.s.}$ of 
Eq.~(\ref{eq:SCHROL1}) is the convolution integral 
\begin{equation} 
 \int \frac{d^2 {\bf q}}{(2\pi)^2} V(\left|{\bf p - q}\right|) 
\phi(\left|{\bf q}\right|) e^{\pm i \ell \theta_q} , 
\label{eq:S1RHS}
\end{equation}
while the F.T of the  $\mbox{l.h.s.}$ of Eq.~(\ref{eq:SCHROL1}) is 
\begin{equation} 
\left\{-\delta + \frac{\left|{\bf p}\right|^2}{m} + 2\mbox{Re}\Sigma_{-}
\right\} \phi(\left|{\bf p}\right|) e^{\pm i \ell \theta_p}
\label{eq:S1LHS}
\end{equation}
Equating Eqs.~(\ref{eq:S1LHS}) and ~(\ref{eq:S1RHS}) we obtain 
\begin{eqnarray} 
\left\{-\delta + \frac{\left|{\bf p}\right|^2}{m} + 
2\mbox{Re}\Sigma_{-} \right\}
 \phi(\left|{\bf p}\right|) & = & 
     \int \frac{d^2 {\bf q}}{(2\pi)^2} 
 V(\left|{\bf p - q}\right|)
 \phi(\left|{\bf q}\right|) e^{\pm i \ell (\theta_q - \theta_p)} 
                       \nonumber \\   
 & = & \int \frac{d^2 {\bf q}}{(2\pi)^2} 
                           V(\left|{\bf p - q}\right|)  
\phi(\left|{\bf q}\right|) \cos \ell(\theta_q - \theta_p) \nonumber \\
 & = & \int \frac{d^2 {\bf q}}{(2\pi)^2}  
               V(\left|{\bf p - q}\right|)
 \phi(\left|{\bf q}\right|) \chi({\bf p},{\bf q}).
 \label{eq:EQUATE1}
\end{eqnarray}
which is equivalent to Eq.~(\ref{eq:OUREQS}).

To summarise, we have obtained equations for degenerate $\ell = 0$ 
vector and axiscalar states (Eq.~(\ref{eq:NRV1})), a new $\ell = 2$ 
vector state (Eq.~(\ref{eq:NRV2})), and degenerate axivector and scalar 
states with $\ell = 1$ (Eq.~(\ref{eq:NRS})).  In order to explain the 
occurrence of this particular set of angular momentum and parity 
combinations we need to study the symmetry properties of bound states 
in (2 + 1) dimensions.  

To classify the bound states in terms of the spin and orbital angular 
momentum parts, we consider briefly the finite dimensional representations 
of the (2 + 1) dimensional Lorentz group $SO(2,1)$ \cite{KN86}.  Its Lie 
algebra is defined by the commutation relations 
\begin{equation}
\left[\Sigma_{\mu},\Sigma_{\nu}\right] =
               i{\epsilon_{\mu \nu}}^\rho \Sigma_\rho, 
\end{equation}
where $\Sigma_0$ is the generator of rotations in the $x_1$-$x_2$ plane 
and $\Sigma_{1,2}$ generate translations in the $x_{1,2}$ directions.  
The finite dimensional representations are classified by 
the Casimir operator 
\begin{equation}
\Sigma^2 = (\Sigma_0)^2 - (\Sigma_1)^2 - (\Sigma_2)^2 , 
\end{equation}
which takes eigenvalues $L(L+1)$, while $\Sigma_0$ takes eigenvalues 
$\ell = -L, -L + 1, \ldots, L - 1, L$.  Identifying $\ell$ with the 
angular dependence of the Schr\"odinger wavefunction discussed above, 
we shall refer to $\ell$ as the total orbital angular momentum.  To the 
spin part of the bound state angular momentum we also attach a quantum 
number $S$, which we call the total spin.  For the symmetric spin 
combinations of two Dirac particles we have $S = 1$, and for the 
antisymmetric combination, $S = 0$.  The spin angular momentum in the 
$x$-$y$ plane takes values $m = -S, -S + 1, \ldots, S - 1, S$.  

The bound states are also classified by their eigenstates under the action 
of the ``axial parity'' operator $A$: 
\begin{equation} 
A: \,\, x^\mu = (x^0,x^1,x^2) \rightarrow x'^\mu = (x^0,-x^1,-x^2), 
                                  \label{eq:AXP}
\end{equation}
which is distinct from the parity operator defining the operation 
Eq.~(\ref{eq:PAR}) in appendix A.  Eq.~(\ref{eq:AXP}) is a rotation 
through $\pi$, and is the operator relevant to interchange of constituent 
fermions.  The transformation properties of Dirac spinors and bilinear 
currents under axial parity are given by Eqs.~(\ref{eq:APAR}) and 
(\ref{eq:APTY}).  The axial parity $A$ of each of the bound states considered 
herein is listed in Table~\ref{tab1}.  

In the non-relativistic limit, the generalised Pauli exclusion principle 
\cite{FS82} adapts from (3+1) to (2+1) dimensions to give the following 
relations between the total orbital angular momentum $L$, the total 
spin $S$, axial parity $A$ and charge parity ${\cal C}$ : 
\begin{equation}
A = (-1)^{L + 1},  \label{al1}
\end{equation}
\begin{equation}
{\cal C} = (-1)^{L + S}.  
\end{equation} 
Two fermions can form a system with a total spin
of $S=0$ or $S=1$.  For a specific total angular momentum $J$, vector
addition of angular momenta $m + \ell$ determines the possible orbital 
angular momenta $\ell$.  Table~\ref{tab1} 
lists all possible scalar and vector states.

Consider the scalar states where $J=0$.  For an orbital angular momentum of
zero ($L=0$) the total spin must also be zero ($S=0$).  In this case
the axial parity must be negative ($A=-1$) and charge parity positive 
(${\cal C}=+1$).  We know from our integral equations that this state is the 
axiscalar state.  It is also known from the transformation properties
outlined in appendix A that the axiscalar does indeed have $A=-1$.  
The other possibility with $J=0$ is the $L=1$, $S=1$ state where the 
axial parity is positive and the charge parity is positive. 
This is the scalar state which we know from its transformation properties 
has $A=+1$.
So we have seen that it is only possible for scalar states to have $L=0$ or
$L=1$ and they must have positive charge parity.
Negative charge parity scalars are forbidden by the generalised Pauli
Exclusion principle.
States such as the $J^{AC}=0^{--}$ state, for example, 
we refer to as having unnatural parity.  

Now consider the vector case ($J=1$).  If $L=0$ then $S$ must equal $1$.  In
this case both the axial parity and charge parity are negative.  
We know that this is the $L=0$ vector state.  With no spin-orbit
coupling contribution in our $e^{-}$--$e^{+}$ potential this state is  
degenerate with the $L=0$ axiscalar state.  

If $L=1$ then there are two possibilities.  A $J=1$ is possible with 
either $S=0$ or $S=1$.  Both of these cases correspond to the
$L=1$ axivector states, both with positive axial parity, 
one with negative charge
parity and the other with positive charge parity.  Again, with no spin-orbit
coupling the two are degenerate.  They are also degenerate
with the $L=1$ scalar state.

The only other possible state is the $L=2$, $S=1$ state which has negative
axial parity and negative charge parity.  This is the $L=2$ vector
state found earlier in this section.
There appears to be no possibility of an $L=2$ vector state with positive
charge parity and so we would hope that there is no such solution in our
nonrelativistic BSE calculations.

In general, we find that the axial parities determined from the angular 
momenta and Eq.~(\ref{al1}) 
match up with those found by analysing the transformation properties 
of bilinear currents in Appendix A.  We also note that the ``axi'' 
transformation in $(2 + 1)$ dimensions parallels the conventional 
parity transformation in $(3 + 1)$ dimensions, and that a one to one 
correspondence between the allowed $J^{AC}$ states in $(2 + 1)$ dimensions 
and $J^{PC}$ states in $(3 + 1)$ dimensions follows from the generalised 
Pauli exclusion principle.  

%------------------------------------------------------------------
\setcounter{equation}{0}
\section{Numerical Results}
In this section we report our numerical solutions to the Bethe-Salpeter 
coupled integral equations of Eq.~(\ref{eq:IE}). 
Comparisons with the nonrelativistic calculations of 
Refs.~\cite{THY95},~\cite{Ko96} will be made for large fermion mass.

As in the scalar calculations of Ref.~\cite{AB96}, a grid of 
$25 \times 25$ ($\mbox{$\left| {\bf q} \right|$}$,$q_3$) 
tiles was used for the iterative solution to the vector equations 
Eq.~(\ref{eq:IE}) with the 
use of linear interpolation on each of those tiles for the sums 
($\tau_{r1}$, $\tau_{r2}$) which are supplied at the corners of the tiles from 
the previous iteration.  Again the tiles were non-uniform in size and 
an upper limit to the momentum components 
($\mbox{$\left| {\bf q} \right|$}$ and $q_3$) were made large enough 
so that results were independent of their values.  
For each symmetry and a range of fermion masses ($0-5$) the equations were
iterated leaving a set of bound state masses.

Table~\ref{tab3} shows the bound 
state masses for each of the four nondegenerate states which are the 
vector ${\cal C}=+1$, vector ${\cal C}=-1$, axivector ${\cal C}=+1$ and the
axivector ${\cal C}=-1$ states.  We employ the notation $J^{AC}$ introduced 
in the last section to classify these states.  
The degeneracy under space reflection parity in 3-dimensions is assumed from
this point on and, for example, the vector state refers to the doublet
vector / pseudovector.
Fig.~1 displays the solutions $M$ for fermion mass 0--0.1.  
Fig.~2 shows $M-2m$ over the greater range of 0--1.   
We note that the poles in the fermion propagator fits lie 
outside the BS integration region $\Omega$ for all solutions obtained.  
This has been the case in both scalar and vector calculations. 

For small $m$ the bound state masses rise rapidly with increasing fermion 
mass.  The $J^{AC} = 1^{--}$ state is the lowest 
energy state followed by the $1^{-+}$ and then the 
$1^{++}$ and $1^{+-}$ states respectively.  
At some point between $m=0.049$ and $0.064$ the $1^{-+}$ and 
the $1^{++}$ states cross. The curves then level off for masses 
approaching the nonrelativistic limit $m>1$.   

We found that the two axivector states were near degenerate for small $m$.
However, numerical problems prevented negative charge parity
solutions to the axivector equations for $m=0.036$ and higher.
The eigenvalue $\Lambda(M)$ in Eq.~(\ref{eq:EV}) splits into complex 
conjugate pairs past this value of $m$.  
We see no theoretical reason for the disappearance of the negative
charge parity axivector state and we attribute it to deficiencies
of the bare vertex, ladder approximation.  

As expected, for larger fermion masses the bound state mass rises 
predominantly as twice the fermion mass plus possible logarithmic 
corrections.  However, there appears to be a good deal of noise in 
the large $m$ solutions, 
reflecting the difficulty in accurately modelling the fermion 
propagator deep into the timelike region from spacelike fits.  
It is for this reason that Figure 2 does not display the larger
fermion masses.  Despite this noise, the ordering of the states remains 
unchanged as $m$ becomes large.

It is clear that this approach to the large fermion mass end of the
spectrum is not suitable and that a special nonrelativistic treatment
is required.  What is required is solution to Eqs.~(\ref{eq:NRV1}), 
~(\ref{eq:NRV2}) and ~(\ref{eq:NRS}).  However, this involves the
solution for the rainbow self energy $\Sigma_{+}$ from a DSE
formulated in the large fermion mass limit.  This is not within the
scope of this paper and is the subject of future work.  In previous
work for the scalar positronium solutions~\cite{AB96} a 1-loop
approximation to this self energy was employed.  This appeared
warranted because the approximation gave a good fit to the rainbow 
propagator for spacelike momenta.  However, it was found that the
1-loop approximation resulted in bound state masses that were out
by a logarithmic correction (because of the noncancelling infrared 
divergences) and as a result the masses could not be compared to
other large fermion mass solutions.  In this work, we do not attempt 
to match 1-loop nonrelativistic solutions
with our large $m$ relativistic solutions.  Despite this, we believe
that the existence or nonexistence of a solution to the 1-loop
nonrelativistic BSE equations is evidence of the existence or
nonexistence of those positronium bound states.  Therefore, 
Eqs.~(\ref{eq:NRV1}),~(\ref{eq:NRV2}) and ~(\ref{eq:NRS}) were
solved with $\Sigma_{+}$ replaced by the 1-loop result in
Eq.~(\ref{eq:SG1LP}) neglecting terms of order $\frac{1}{m^2}$.

Because of the degeneracies pointed out in section III,
solutions exist to Eqs.~(\ref{eq:NRV1}) and ~(\ref{eq:NRS}) and from
this we could conclude that the vector $\ell=0$ ($1^{--}$) and the 
axivector $\ell=1$ ($1^{-\pm}$) bound states do exist.  
We were left to solve Eq.~(\ref{eq:NRV2}) for
possible positive and negative charge parity $\ell=2$ vector solutions.
It was found that the $1^{--}$ solution did exist and had a mass
increasing linearly in $\ln m$ as did the smaller $\ell$ solutions.
However, the $1^{-+}$ solution mass did not level off to the same 
slope and appeared to have a strongly divergent mass in the nonrelativistic 
limit.  We interpret this by saying that there is no $1^{-+}$ bound state,
in agreement with the generalised Pauli exclusion principle.

Although the 1-loop nonrelativistic masses cannot be quantitatively compared 
with masses found with the rainbow propagator, we again point out that the
results do contain useful qualitative information.  We find that 
the ordering of the states in the nonrelativistic limit is
$1^{--}$ ($\ell=0$) followed by $1^{+\pm}$ ($\ell=1$) and then 
$1^{--}$ ($\ell=2$).  
Based on both the  relativistic calculations and the nonrelativistic 
1-loop exercise, it appears that it is energetically favourable for the 
scalars to have positive charge parity, and for the vectors to have negative 
charge parity, in agreement with the QCD$_4$ meson analogy.

We compare our relativistic integral equation results with those of existing 
QED$_3$ Schr\"odinger equation results.  
Tam, Hamer and Yung \cite{THY95} perform an analysis of QED$_3$ from the point of 
view of discrete light cone quantisation.  In the non-relativistic limit, 
their bound state masses $M$ are solution to the Schr\"odinger equation 
\begin{equation} 
\left\{ -\frac{1}{m} \nabla^2 + \frac{1}{2\pi} \left(C + \ln(mr) 
 \right)\right\}  \phi({\bf r}) = (M - 2m) \phi({\bf r}), \label{eq:THYDE}     
\end{equation}
where $C$ is Euler's constant and $m$ is the bare fermion mass.
This differential equation is solved for the completely symmetric states
and the masses are given by the expression 
\begin{equation}
M  =  2m + \frac{1}{4\pi} \ln m  + \frac{1}{2\pi} \left(\lambda^{'} -
\frac{1}{2} \ln\frac{2}{\pi} \right).  
       \label{eq:THY1}
\end{equation}
Koures \cite{Ko96} solves the same equation but also for nonzero angular
momentum $\ell$.  For $0 \le \ell \le 4$ this reference provides the 
lowest five eigenvalues ($\lambda$) which, after the transformation
\begin{equation}
\lambda^{'}=\lambda + \ln(2) + C
\label{eq:LPRIME}
\end{equation}
can be used in Eq.~(\ref{eq:THY1}) above.  The eigenvalues to be used here are 
$\lambda' = 1.7969$ ($\ell = 0$), 2.6566 ($\ell = 1$), 2.9316 ($\ell = 0$) and 
3.1148 ($\ell = 2$).  

The lowest $\ell=$ 0, 1 and 2  and the first excited $\ell=$ 0 Schr\"odinger 
equation results of 
Refs.~\cite{THY95},~\cite{Ko96} are also plotted in Fig.~2. 
One could associate the $\ell=$ 0 curve with the vector $1^{--}$
state.  The $\ell=$ 1 curve can be seen to match up with one or more 
of $1^{-+}$, $1^{++}$ or $1^{+-}$.  We know from our nonrelativistic
analysis that the $1^{+\pm}$ (axivector) states are the only $\ell=$ 1 
states and become degenerate in the large $m$ limit.  It therefore seems
as though there is a surprisingly good match up between relativistic and 
nonrelativistic $\ell=$ 1 states.  
The lowest $\ell=$ 2 and the first excited $\ell=0$ curves should coincide
with higher vector $1^{--}$ states
but no such states were solved for in our relativistic BSE treatment.
Also there is no nonrelativistic $1^{-+}$ state to match up with the
relativistic vector state with positive charge parity as shown in
our nonrelativistic BSE exercise and as predicted by the generalised
Pauli exclusion principle.
We emphasise that the Schr{\" o}dinger equation results are of
course formulated in the nonrelativistic limit and we cannot expect close
numerical agreement with the small mass relativistic results.  In fact,
there is quite a broad region of intermediate masses where neither the
relativistic BSE nor the Schr{\" o}dinger results are suitable.

Finally, we may make qualitative comparisons between the calculated spectrum 
of $J^{AC}$ states in QED$_3$ and the spectrum of observed $J^{PC}$ meson 
states in QCD$_4$.  Since the observed meson spectrum is best established 
for light mesons, the comparison is made in the chiral limit $m=0$.  
In Table~\ref{tabmes} we list the $m=0$ spectrum of scalar and axiscalar 
states from Ref.~\cite{AB96}, and of vector and axivector states from this 
work, together with the corresponding observed light meson states 
~\cite{PRG96}.  (Note 
that, as pointed out in the previous section, the axial parity quantum 
number $A$ in $(2 + 1)$ dimensions plays the role of conventional parity 
$P$ in $(3 + 1)$ dimensions.)  Firstly we see that the gaps in the observed 
meson spectrum occur for parity combinations disallowed by the generalised 
Pauli exclusion principle.  In our fully relativistic QED$_3$ calculations, 
there is nothing in principle to prevent such states from occurring away from 
the nonrelativistic limit.  We are unable to say whether the unnatural parity 
solutions are a genuine property of QED$_3$, or an artefact of the 
rainbow, ladder approximation.  If these states are ignored, we observe 
surprising agreement between the ordering and relative magnitudes of 
bound state masses in QED$_3$ and the observed meson spectrum.  
In both cases the dynamics of the bound state spectrum is driven mainly by 
confinement and chiral symmetry breaking, these being the common features 
of QED$_3$ and QCD$_4$.  

%------------------------------------------------------------------
\setcounter{equation}{0}
\section{Conclusions} 

We have extended our earlier study of the positronium states of 
4-component QED$_3$ from scalar to vector states.  QED$_3$ is a confining 
theory, and the positronium states are in some sense the analogues of mesons 
in QCD$_4$.  In $(2 + 1)$ dimensions the bound states are classified as 
eigenstates 
of reflection parity $P$, axial parity $A$ (i.e. a spatial rotation 
through $180^\circ$), and charge conjugation $C$.  In the four component 
version of massless QED$_3$, a $U(2)$ symmetry analogous to chiral symmetry is 
spontaneously broken to $U(1)\times U(1)$.  The resulting spectrum consists 
of reflection parity doublets for which $P$ is an exact symmetry.  The 
states are therefore classified by the quantum numbers $J^{AC}$.  We 
had previously studied in detail axiscalar ($0^-$) and scalar ($0^+$) 
positronium.  Herein our focus is mainly on the extension of this work 
to vector ($1^-$) and axivector ($1^+$) states.  

We calculate bound state masses using the combination of rainbow 
Dyson-Schwinger and ladder Bethe-Salpeter equations.  Fermion propagators 
calculated for spacelike momenta from the Dyson-Schwinger equation are 
extended into the required part of the complex momentum plane by making 
analytic fits to the spacelike part of the propagator.  There are two 
important issues raised by this extrapolation procedure, both of which 
we addressed for the scalar case in our previous work.  

Firstly, the solution to the rainbow approximation Dyson-Schwinger 
equation, and the analytic fits, contain ghost poles in 
the timelike half of the complex momentum plane.  It is necessary for the 
success of the approximation that these poles do not impinge on the set of 
complex momentum values sampled by the BSE.  As for the scalar case 
considered previously, we find that the vector states do not pose a problem 
in this regard.  This is because the scalar and vector bound state masses 
are typically of comparable size over the range of bare fermion masses 
considered.  

Secondly, in both the scalar and vector cases, the extrapolation procedure 
is inadequate for intermediate or large bare fermion masses 
(viz. $m/e^2 > 0.5$).  This is because the important contributions 
to the BSE in the heavy fermion limit come from deep into the timelike 
part of the complex plane.  For this reason the heavy fermion limit must 
be treated separately.  

We have obtained numerical solutions to the combination of Dyson-Schwinger 
and Bethe-Salpeter equations over a range of dimensionless bare fermion 
masses $m/e^2$ for each of the axial and charge parity combinations 
$1^{--}$, $1^{-+}$, $1^{+-}$ and $1^{++}$.  Solutions exist over the 
broad range of bare fermion masses considered except for the $1^{+-}$, 
for which the eigenvalue of the integral operator in the BSE becomes 
complex for $m/e^2 > 0.036$.  We interpret this as a shortcoming of the 
rainbow-ladder approximation.  

In our previous work we obtained the nonrelativistic 
(i.e. $m\rightarrow \infty$) limit of the 
BSE for scalar and axiscalar states in the form of a Schr\"odinger equation.  
Here we have extended the proof to vector and axivector states in a 
way which enables us to identify the orbital angular momentum of each 
state.  We find that the relation $A = (-1)^{L+1}$ is automatically 
satisfied, and are furthermore 
able to restrict the allowed charge parities by using 
the generalised Pauli exclusion principle.  In this limit it becomes clear 
that it is axial parity, and not space reflection parity, which is the 
QED$_3$ counterpart of conventional parity in QCD$_4$.  

We also find that in the nonrelativistic limit, the bound state 
spectrum becomes insensitive to the spin of constituent fermions.  This 
is manifested as a degeneracy between $0^{-+}$ and $1^{--}$ states 
and between the $0^{++}$ and $1^{+\pm}$ states.  Precisely the same 
phenomenon occurs in the observed heavy meson spectrum to within order 
of the inverse heavy quark mass, and can be explained 
in terms of heavy quark effective theory~\cite{N94}.  

While certain $J^{AC}$ combinations are disallowed by the generalised 
Pauli exclusion principle in the nonrelativistic limit, there is nothing 
in principle to prevent their occurrence in the relativistic regime.  
We have in fact obtained unnatural parity solutions in both our earlier 
scalar and current vector positronium calculations.  This is not in agreement 
with the observed meson spectrum, even for light quark mesons, where 
for instance negative charge parity scalar or pseudoscalar mesons are never 
observed.  One is led to question whether the unnatural parity solutions in 
QED$_3$ are an artefact of the rainbow-ladder approximation.  More importantly 
we find that, if the unnatural parity solutions are ignored, 
there is surprising agreement between the relative 
mass scales of calculated $J^{AC}$ states in QED$_3$ and observed $J^{PC}$ 
mesons in QCD$_4$, as evidenced in Table~\ref{tabmes}. 

The original aim of studying 4-component QED$_3$ was to explore 
nonperturbatively a theory which has properties in common with QCD$_4$, 
namely confinement and chiral symmetry breaking, but without the 
complications of being nonabelian.  We have demonstrated that the 
Bethe-Salpeter formalism generates a bound state spectrum with qualitative 
characteristics in common with the observed meson spectrum.  
One shortcoming of the rainbow-ladder formalism is that gauge covariance
is broken by using a bare fermion-photon vertex.  In any practical
truncation it is likely that some symmetry of the original problem will
be lost.  In this paper we have adopted the position that it is more
important for bound state mass calculations to maintain the chiral
symmetry breaking mechanism than to maintain gauge symmetry.  An alternative
approach is that of lattice gauge theory, in which gauge symmetry is 
maintained, but the original chiral symmetry of QED$_3$ survives only as a
small remnant subgroup~\cite{BB87}.  
Specifically, lattice gauge theory simulations of quenched, non-compact
QED3 with 4-component fermions can be carried out with comparative
ease~\cite{DKK90}, and the chiral condensate checked against the predictions
of model Dyson-Schwinger calculations~\cite{B92}.  We are unaware of any
existing bound state spectrum calculation of this simple lattice model.
Such a calculation would provide a much needed cross check between two vastly
differing non-perturbative field theory methods.

%------------------------------------------------------------------
\renewcommand{\theequation}{\Alph{section}.\arabic{equation}}
\setcounter{section}{1}
\setcounter{equation}{0}
\section*{Appendix A - General Vector Amplitude in QED$_3$}
Solution to the BSE in Eq.~(\ref{eq:BS}) for the vector states requires 
a general vertex function which will allow individual Dirac components to be
projected out leaving a system of coupled integral equations.  We assume
a knowledge of the transformation properties of QED$_3$~\cite{Bu92} and of the 
Scalar BS amplitudes~\cite{AB96}. 

We work in the four-component version of QED$_3$ where there exists a
complete set of 16 matrices (where $\mu$ = 0, 1 and 2)
$\{\gamma_A\}=\{I,\gamma_{4},\gamma_{5},\gamma_{45},\gamma_{\mu},
         \gamma_{\mu 4},\gamma_{\mu 5},\gamma_{\mu 45} \}$
satisfying $\frac{1}{4} {\rm tr}(\gamma_A \gamma^B) = \delta^B_A$.
These matrices can be found in Ref.~\cite{Bu92}. 
The QED$_3$ action is invariant with respect to discrete parity and 
charge conjugation symmetries, which for the fermion fields are
given by
\begin{equation}
\mbox{$\psi(x)$} \rightarrow \psi^\prime(x^\prime) = 
\Pi \mbox{$\psi(x)$}, \;\;\;
\mbox{$\overline{\psi}(x)$} \rightarrow \overline{\psi}^\prime
(x^\prime) = \mbox{$\overline{\psi}(x)$} \Pi^{-1},
                  \label{eq:PAR}
\end{equation}
\begin{equation}
\mbox{$\psi(x)$} \rightarrow \psi^\prime(x) = 
C \overline{\psi}(x)^{\rm T}, \;\;\;
\mbox{$\overline{\psi}(x)$} \rightarrow \overline{\psi}^\prime(x) 
= -\mbox{$\psi(x)$}^{\rm T} C^{\dagger},
                    \label{eq:CH}
\end{equation}
where $x^{\prime}=(x^0,-x^1,x^2)$. The parity and charge conjugation
matrices ($\Pi$ and $C$ respectively) are
each determined up to an arbitrary phase by the condition that
the action be invariant~\cite{Bu92}:
\begin{equation}
\Pi=\gamma_{14} e^{i\phi_P \gamma_{45}}, \;\;\;
 C=\gamma_{2} e^{i\phi_C \gamma_{45}}, (0\leq \phi_P,\phi_C < 2\pi)
\label{eq:pic} 
\end{equation}

Vector, pseudovector, axivector and axipseudovector bound states are
defined by the following transformation properties under
parity transformations 
\begin{eqnarray}
\Phi_{\mu}^{V}(x) & \rightarrow & 
          \Phi_{\mu}^{V\prime}(x^{\prime}) = 
         \Lambda^{\nu}_{\mu} \Phi_{\nu}^{V}(x),\nonumber \\
\Phi_{\mu}^{PV}(x) & \rightarrow & \Phi_{\mu}^{PV\prime}(x^{\prime}) =
        -\Lambda^{\nu}_{\mu} \Phi_{\nu}^{PV}(x), \nonumber \\
\Phi_{\mu}^{AV}(x) & \rightarrow & \Phi_{\mu}^{AV\prime}(x^{\prime}) = 
         R_P \Lambda^{\nu}_{\mu} \Phi_{\nu}^{AV}(x),\nonumber \\
\Phi_{\mu}^{APV}(x) & \rightarrow & \Phi_{\mu}^{APV\prime}(x^{\prime}) = 
        -R_P \Lambda^{\nu}_{\mu} \Phi_{\nu}^{APV}(x), \label{eq:PTY} 
\end{eqnarray}
Where
\begin{eqnarray}
R_P & = & \left(\begin{array}{cc} -\cos 2\phi_P & -\sin 2\phi_P \\
        -\sin 2\phi_P & \cos 2\phi_P  \end{array}  \right)
\label{eq:RL}
\end{eqnarray}
and in Minkowski space $\Lambda_{\mu}^{\nu}={\rm diag}(1,-1,1)$.
Similar transformation properties exist for charge conjugation.

We also find the need to classify our nonrelativistic states in terms
of the transformation properties under what we call ``axial parity''
which is the (2+1)d analogue of the (3+1)d parity where
$x^{\prime}=(x^0,-x^1,-x^2)$.  The fermion fields transform like
\begin{equation}
\mbox{$\psi(x)$} \rightarrow \psi^\prime(x^\prime) = 
S_{\pi} \mbox{$\psi(x)$}, \;\;\;\;\;\;
\mbox{$\overline{\psi}(x)$} \rightarrow \overline{\psi}^\prime
(x^\prime) = \mbox{$\overline{\psi}(x)$} S_{\pi}^{-1},
                  \label{eq:APAR}
\end{equation}
where $S_{\pi}$ is the operator which corresponds to a rotation through
angle $\pi$ in the $x^1x^2$ plane.  We find that a suitable
operator $S_{\pi}$ which performs such a rotation and leaves the action
invariant is the matrix $i \gamma_0$.  This is the same operator as used
in the (3+1)d case.  The phase $i$ is responsible for making 
$S_{\pi}^{2}=-1$ and 
has no effect on the transformation of bilinear currents of concern here.  
Given that the bound state wavefunctions transform like
the bilinear currents $J_A=\overline{\psi}(x) \gamma_A \psi(x)$ 
we find the following transformation properties for scalars and 
vectors under axial parity.
\begin{eqnarray}
  \Phi^{S}(x)   \rightarrow&  \,\,\,\,\Phi^{S}(x),    \,\,\,\,\,\,\,\,\,\,\,\,\,\,\,\,\,\,\,\,\,\, 
 \Phi_{\mu}^{V}(x)  &\rightarrow \,\,\,\, \eta^{\nu}_{\mu} \Phi_{\nu}^{V}(x),  \nonumber \\
  \Phi^{PS}(x)  \rightarrow& \,\,\,\, \Phi^{PS}(x),     \,\,\,\,\,\,\,\,\,\,\,\,\,\,\,\,\,\,\,\,\,\, 
 \Phi_{\mu}^{PV}(x)    &\rightarrow  \,\,\,\, \eta^{\nu}_{\mu} \Phi_{\nu}^{PV}(x),   \nonumber \\
  \Phi^{AS}(x)  \rightarrow& -\Phi^{AS}(x), \,\,\,\,\,\,\,\,\,\,\,\,\,\,\,\,\,
 \Phi_{\mu}^{AV}(x)    &\rightarrow  - \eta^{\nu}_{\mu} \Phi_{\nu}^{AV}(x),   \nonumber \\ 
  \Phi^{APS}(x)  \rightarrow&  -\Phi^{APS}(x),   \,\,\,\,\,\,\,\,\,\,\,\,\,\,\,\,\, 
 \Phi_{\mu}^{APV}(x)   &\rightarrow  - \eta^{\nu}_{\mu} \Phi_{\nu}^{APV}(x). \label{eq:APTY} 
\end{eqnarray}

The matrix $\eta_{\mu}^{\nu}$ is defined as ${\rm diag}(1,-1,-1)$.  
The corresponding axial parities ($A=\pm 1$) for each state are 
listed in Table~\ref{tab1}.  
This definition parallels the (3+1)d case where a vector state transforms
like $\eta^{\nu}_{\mu} \Phi_{\nu}$ and is assigned negative parity.

Let us now derive a general vector vertex.  Consider the momentum $p_{\mu}$
which transforms like $p_{\mu} \rightarrow \Lambda_{\mu}^{\nu} p_{\nu}$
under parity.  The bound state momentum $P_{\mu}$ seen in Eq.~(\ref{eq:BS})
is another such vector.
Three other vectors can be found which transform in this way which are
$\gamma_{\mu}$, $u_{\mu}(p)$ and $v_{\mu}(p)\gamma_{45}$.  The vectors
$u(p)$ and $v(p)$ used in previous work~\cite{Bu92} are mutually
orthogonal with $P$ and are defined as
\begin{equation}
u(p)=\frac{P^2p-(P.p)P}{P^2p^2-(P.p)^2}  \;\;\;\;\; , \;\;\;\;\;
v(p)=\frac{P \times p}{P^2p^2-(P.p)^2}.
\end{equation}
The proof that these vectors transform like $p$
is simple using the knowledge that a vector $V_{\mu}(p,P)$ transforms under parity like
$V_{\mu}(p,P) \rightarrow \Pi V_{\mu}(\Lambda p, \Lambda P) \Pi^{-1} $
and that the five matrices $\gamma_{\mu}$, 
$\gamma_{4}$ and $\gamma_{5}$ form an anticommuting set and 
$\gamma_{45}=-i \gamma_{4}\gamma_{5}$.

So we have a set of five vectors with even parity.   A linear
combination of $P$ and $u$ can be found which gives $p$ and so it is not
required.  The four vectors we work with are thus $P_{\mu}$, $\gamma_{\mu}$, 
$u_{\mu}(p)$ and $v_{\mu}(p)\gamma_{45}$.  Any other available vectors even 
in parity would be linear combinations of these.

A massive boson of spin 1 will have a purely transverse wavefunction
~\cite{LS69}.
Thus the vector vertex will be purely transverse.  This
zero divergence criterion in momentum space means that 
$P_{\mu} \Gamma_{\mu} = 0$.  
We must find linear combinations of the four vectors above which are all
purely transverse.  Only three can be found which are
$T_{\mu}(P)=\gamma_{\mu}-P_{\mu} \not \! P / P^2$, 
$u_{\mu}(p)$ and $v_{\mu}(p)\gamma_{45}$.  It therefore appears as though 
the most general vector $\Gamma_{\mu}(p,P)$ we can write which is even under 
parity and purely transverse is a sum of these quantities, each multiplied 
by the most general scalar vertex~\cite{Bu92}.  The result is an
expression with twelve terms and with twelve parameters.
However, some terms have been included more than once.  It can be shown 
that four of the twelve terms can be eliminated leaving the general vector 
vertex
\begin{eqnarray}
\Gamma^V_{\mu}(p,P) & = & u_{\mu}(p)   \left( f_1 + f_2 \not \! P + 
f_3 \not \! u(p) + f_4 \not \! v(p)\gamma_{45} \right) \nonumber \\
                  & + & v_{\mu}(p)\gamma_{45}  \left( f_5 + f_6 \not \! P + 
f_7 \not \! u(p) + f_8 \not \! v(p)\gamma_{45} \right). \label{eq:VERTEX}
\end{eqnarray}
where $f_n$ ($n=$1--8) are functions of $q^2,P^2$ and $q\cdot P$ only.   
The pseudovector (PV), axivector (AV) and axipseudovector (APV) vertices 
are given by
\begin{eqnarray}
\Gamma^{PV}_{\mu}(p,P) & = &   \gamma_{45} \Gamma^V_{\mu}(p,P), \nonumber \\
\Gamma^{AV}_{\mu}(p,P) & = & \left(\begin{array}{c} \gamma_4 \\ 
      \gamma_5 \end{array} \right) \Gamma^V_{\mu}(p,P), \nonumber \\
\Gamma^{APV}_{\mu}(p,P) & = &   \gamma_{45} \Gamma^{AV}_{\mu}(p,P) \,\,\,  =  
\,\,\,  i \left(\begin{array}{c} -\gamma_5 \\ 
      \gamma_4 \end{array} \right) \Gamma^V_{\mu}(p,P). \label{eq:AVS}
\end{eqnarray}

For a specified charge parity ${\cal C}=\pm 1$ of a bound state, the parity 
of the Dirac coefficients ($f_n$) under the transformation
$q\cdot P \rightarrow -q\cdot P$ can be determined.  The quantity 
$q\cdot P$ is the only Lorentz invariant which changes sign under charge 
conjugation and thus 
determines the charge parity of those functions.
Table~\ref{tab5} lists the charge parities of each function for each case.

We can see from this table that multiplying the vector vertex by $\gamma_{45}$ 
(and thus producing the pseudovector vertex) leaves the function charge 
parities unchanged but multiplying the axivector vertex by $\gamma_{45}$ 
(producing the axipseudovector vertex) reverses the function charge
parities.  
We know that the same BSE results when we multiply the vertex
$\Gamma_\mu$ by $\gamma_{45}$. 
This means that the vector ${\cal C}=+1$ and pseudovector ${\cal C}=+1$
states are degenerate as are the vector ${\cal C}=-1$ and pseudovector 
${\cal C}=-1$ states.  We must also find the degenerate pairs
axivector ${\cal C}=+1$ / axipseudovector ${\cal C}=-1$ and
axivector ${\cal C}=-1$ / axipseudovector ${\cal C}=+1$.   

Our conventions for Euclidean space quantities are summarised in
Appendix A of Ref.~\cite{Bu92}.  In particular Euclidean momenta and 
Dirac matrices are defined by
$$
P_3^{({\rm E})} = -iP_0^{({\rm M})},\hspace{5 mm}
P_{1,2}^{({\rm E})} = P_{1,2}^{({\rm M})},\hspace{5 mm}
\gamma_3^{({\rm E})} = \gamma_0^{({\rm M})}, \hspace{5 mm}
\gamma_{1,2}^{({\rm E})} = i\gamma_{1,2}^{({\rm M})}.
$$

%-----------------------------------------------------------------
\renewcommand{\theequation}{\Alph{section}.\arabic{equation}}
\setcounter{section}{2}
\setcounter{equation}{0}
\section*{Appendix B - Coupled Integral Equations}

Working in the rest frame of the bound state where $P_{\mu}=(0,0,iM)$,
the vector and axivector vertices given in appendix A are used
to derive coupled BS equations.  After considerable work the eight
coupled equations were found to be (after rescaling and angular
integration)
\begin{eqnarray}
a(p) & = & \frac{3}{(2\pi)^2} \int^{\infty}_{-\infty}dq_3 \,
     \int^{\infty}_0 \mbox{$\left| {\bf q} \right|$} d\mbox{$\left| 
     {\bf q} \right|$} \, X(\alpha,\beta) \,\, \tau_{a1}(q) \nonumber \\
b(p) & = & \frac{1}{(2\pi)^2} \int^{\infty}_{-\infty}dq_3 \,
  \int^{\infty}_0 \mbox{$\left| {\bf q} \right|$} d\mbox{$\left| 
  {\bf q} \right|$} \,  X(\alpha,\beta) \,\, \tau_{b1}(q)    \nonumber \\    
c(p) & = & \frac{1}{(2\pi)^2} \int^{\infty}_{-\infty}dq_3 \, \int^{\infty}_0 
\mbox{$\left| {\bf q} \right|$} d\mbox{$\left| {\bf q} \right|$} \,  
\left[ Z(\alpha,\beta) \,\, \tau_{c1}(q) + Y(\alpha,\beta) \,\, \tau_{c2}(q) 
\right] \nonumber \\  
d(p) & = & \frac{1}{(2\pi)^2} \int^{\infty}_{-\infty}dq_3 \, \int^{\infty}_0 
\mbox{$\left| {\bf q} \right|$} d\mbox{$\left| {\bf q} \right|$} \,  
\left[ Z(\alpha,\beta) \,\, \tau_{d1}(q) + Y(\alpha,\beta) \,\, \tau_{d2}(q) 
\right] \nonumber \\  
e(p) & = & \frac{3}{(2\pi)^2} \int^{\infty}_{-\infty}dq_3 \, \int^{\infty}_0 
\mbox{$\left| {\bf q} \right|$} d\mbox{$\left| {\bf q} \right|$} \, 
     X(\alpha,\beta) \,\, \tau_{e2}(q) \nonumber \\
f(p) & = & \frac{1}{(2\pi)^2} \int^{\infty}_{-\infty}dq_3 \, \int^{\infty}_0 
\mbox{$\left| {\bf q} \right|$} d\mbox{$\left| {\bf q} \right|$} \, 
     X(\alpha,\beta) \,\, \tau_{f2}(q) \nonumber \\
g(p) & = & \frac{1}{(2\pi)^2} \int^{\infty}_{-\infty}dq_3 \, \int^{\infty}_0 
\mbox{$\left| {\bf q} \right|$} d\mbox{$\left| {\bf q} \right|$} \,  
\left[ Y(\alpha,\beta) \,\, \tau_{g1}(q) + Z(\alpha,\beta) \,\, \tau_{g2}(q) 
\right] \nonumber \\  
h(p) & = & \frac{1}{(2\pi)^2} \int^{\infty}_{-\infty}dq_3 \, \int^{\infty}_0 
\mbox{$\left| {\bf q} \right|$} d\mbox{$\left| {\bf q} \right|$} \,  
\left[ Y(\alpha,\beta) \,\, \tau_{h1}(q) + Z(\alpha,\beta) \,\, \tau_{h2}(q) 
\right] \label{eq:IE}  
\end{eqnarray}
where we have rescaled the functions $f_1,f_2,f_3,f_4,f_5,f_6,f_7$ and $f_8$
seen in appendix A to the new functions $a,b,c,d,e,f,g$ and $h$
to ensure all quantities are real.  The rescalings for the vector case were
\begin{eqnarray}
\vspace{2mm}
a(p_3,\mbox{$\left| {\bf p} \right|$})&=
\frac{1}{\mbox{$\left| {\bf p} \right|$}} \, 
f_1(p_3,\mbox{$\left| {\bf p} \right|$}), \,\,\,\,\,\,\,\,\,\,\,\,\,\,\,\,    
e(p_3,\mbox{$\left| {\bf p} \right|$})&=
\frac{1}{M \mbox{$\left| {\bf p} \right|$}} \,  
f_5(p_3,\mbox{$\left| {\bf p} \right|$}), \nonumber \\ \vspace{2mm}
b(p_3,\mbox{$\left| {\bf p} \right|$})&=
\frac{M}{\mbox{$\left| {\bf p} \right|$}} \, 
f_2(p_3,\mbox{$\left| {\bf p} \right|$}), \,\,\,\,\,\,\,\,\,\,\,\,\,\,\,\,    
f(p_3,\mbox{$\left| {\bf p} \right|$})&=
\frac{1}{\mbox{$\left| {\bf p} \right|$}} \, 
f_6(p_3,\mbox{$\left| {\bf p} \right|$}),\nonumber \\  \vspace{2mm}
c(p_3,\mbox{$\left| {\bf p} \right|$})&=
\frac{i}{\mbox{$\left| {\bf p} \right|$}^2} \, 
f_3(p_3,\mbox{$\left| {\bf p} \right|$}), \,\,\,\,\,\,\,\,\,\,\,\,\,\,\,\,    
g(p_3,\mbox{$\left| {\bf p} \right|$})&=
\frac{-i}{M \mbox{$\left| {\bf p} \right|$}^2} \, 
f_7(p_3,\mbox{$\left| {\bf p} \right|$}), \nonumber \\   \vspace{2mm}
d(p_3,\mbox{$\left| {\bf p} \right|$})&=
\frac{-i}{M \mbox{$\left| {\bf p} \right|$}^2} \, 
f_4(p_3,\mbox{$\left| {\bf p} \right|$}), \,\,\,\,\,\,\,\,\,\,\,\,\,\,\,\,    
h(p_3,\mbox{$\left| {\bf p} \right|$})&=
\frac{i}{M^2 \mbox{$\left| {\bf p} \right|$}^2} \, 
f_8(p_3,\mbox{$\left| {\bf p} \right|$}), 
\label{eq:RESV}
\end{eqnarray}
and for the axivector case
\begin{eqnarray}
\vspace{2mm}
a(p_3,\mbox{$\left| {\bf p} \right|$})&=
\frac{1}{\mbox{$\left| {\bf p} \right|$}} \, 
f_1(p_3,\mbox{$\left| {\bf p} \right|$}), \,\,\,\,\,\,\,\,\,\,\,\,\,\,\,\, 
e(p_3,\mbox{$\left| {\bf p} \right|$})&=
\frac{i}{M \mbox{$\left| {\bf p} \right|$}} \, 
f_5(p_3,\mbox{$\left| {\bf p} \right|$}),  \nonumber \\  \vspace{2mm}
b(p_3,\mbox{$\left| {\bf p} \right|$})&=
\frac{-i M}{\mbox{$\left| {\bf p} \right|$}} \, 
f_2(p_3,\mbox{$\left| {\bf p} \right|$}), \,\,\,\,\,\,\,\,\,\,\,\,\,\,\,\, 
f(p_3,\mbox{$\left| {\bf p} \right|$})&=
\frac{1}{\mbox{$\left| {\bf p} \right|$}} \, 
f_6(p_3,\mbox{$\left| {\bf p} \right|$}), \nonumber \\  \vspace{2mm}
c(p_3,\mbox{$\left| {\bf p} \right|$})&=
\frac{1}{\mbox{$\left| {\bf p} \right|$}^2} \, 
f_3(p_3,\mbox{$\left| {\bf p} \right|$}), \,\,\,\,\,\,\,\,\,\,\,\,\,\,\,\, 
g(p_3,\mbox{$\left| {\bf p} \right|$})&=
\frac{-i}{M \mbox{$\left| {\bf p} \right|$}^2} \, 
f_7(p_3,\mbox{$\left| {\bf p} \right|$}), \nonumber \\  \vspace{2mm}
d(p_3,\mbox{$\left| {\bf p} \right|$})&=
\frac{-i}{M \mbox{$\left| {\bf p} \right|$}^2} \, 
f_4(p_3,\mbox{$\left| {\bf p} \right|$}), \,\,\,\,\,\,\,\,\,\,\,\,\,\,\,\, 
h(p_3,\mbox{$\left| {\bf p} \right|$})&=
\frac{1}{M^2 \mbox{$\left| {\bf p} \right|$}^2} \, 
f_8(p_3,\mbox{$\left| {\bf p} \right|$}). 
\label{eq:RESAV}
\end{eqnarray}
These functions also occur within the sums ($\tau$) seen in 
Eq.~(\ref{eq:IE}) which are defined as (for a function $r=a$--$h$)
\begin{eqnarray}
\tau_{r1}(q)&=&T_{ra}\,a(q)+T_{rb}\,b(q)+T_{rc}\,c(q)+T_{rd}\,d(q) 
\,\,\,\,\,\, \mbox{and} \nonumber \\ 
\tau_{r2}(q)&=&T_{re}\,e(q)+T_{rf}\,f(q)+T_{rg}\,g(q)+T_{rh}\,h(q).
\end{eqnarray}
Also
$$
X(\alpha,\beta)=\frac{(\alpha^2-\beta^2)^{\frac{1}{2}}-\alpha}{\beta
(\alpha^2-\beta^2)^{\frac{1}{2}}}, \,\,\,\,\,\,\,\,\,\,\,\,\,
Y(\alpha,\beta)=\frac{1}{\alpha+(\alpha^2-\beta^2)^{\frac{1}{2}}}, 
$$
\begin{equation}
Z(\alpha,\beta)=\frac{\alpha}{\alpha(\alpha^2-\beta^2)^{\frac{1}{2}}+
(\alpha^2-\beta^2)},
\label{eq:XYZdef}
\end{equation}
where
$$
\alpha=(p_3-q_3)^2 + \mbox{$\left| {\bf p} \right|$}^2 + \mbox{
$\left| {\bf q} \right|$}^2, \;\;\;\;\;\;
\beta=-2 \mbox{$\left| {\bf p} \right|$} \mbox{$\left| {\bf q} \right|$},
$$
The momentum Q is defined by
\begin{equation}
Q^2=q_3^2 + \mbox{$\left| {\bf q} \right|$}^2 -\frac{1}{4}M^2 
+ iMq_3 , \label{eq:QDEF}
\end{equation}
and we use the abbreviations $\sigma_V=\sigma_V(Q^2)$ and 
$\sigma_S=\sigma_S(Q^2)$ for use in the definition of the 
functions $T_{aa},T_{ab},\ldots$ which are analytic 
functions of $q_3, \mbox{$\left| {\bf q} \right|$}$, and $M$.  
The diagonal $T$'s are given by 
\begin{eqnarray}
 T_{aa}=\,\,\,\,\,(\frac{1}{4}M^2 + q_3^{\,2} + \mbox{$\left| {\bf q} 
\right|$}^2) |\sigma_V|^2 \mp |\sigma_S|^2, \,\,\,\,\,
 &T_{ee}=\,\,\,\,\,(\frac{1}{4}M^2 + q_3^{\,2} + \mbox{$\left| {\bf q} 
\right|$}^2) |\sigma_V|^2 \pm |\sigma_S|^2 \nonumber \\
 T_{bb}=-(\frac{1}{4}M^2 + q_3^{\,2} - \mbox{$\left| {\bf q} 
\right|$}^2) |\sigma_V|^2 \pm |\sigma_S|^2, \,\,\,\,\,
 &T_{ff}=-(\frac{1}{4}M^2 + q_3^{\,2} - \mbox{$\left| {\bf q} 
\right|$}^2) |\sigma_V|^2 \pm |\sigma_S|^2 \nonumber \\
 T_{cc}=\,\,\,\,\,(\frac{1}{4}M^2 + q_3^{\,2} - \mbox{$\left| {\bf q} 
\right|$}^2) |\sigma_V|^2 \pm |\sigma_S|^2, \,\,\,\,\,
 &T_{gg}=\,\,\,\,\,(\frac{1}{4}M^2 + q_3^{\,2} - \mbox{$\left| {\bf q} 
\right|$}^2) |\sigma_V|^2 \pm |\sigma_S|^2 \nonumber \\
 T_{dd}=\,\,\,\,\,(\frac{1}{4}M^2 + q_3^{\,2} + \mbox{$\left| {\bf q} 
\right|$}^2) |\sigma_V|^2 \pm |\sigma_S|^2, \,\,\,\,\,
 &T_{hh}=\,\,\,\,\,(\frac{1}{4}M^2 + q_3^{\,2} + \mbox{$\left| {\bf q} 
\right|$}^2) |\sigma_V|^2 \pm |\sigma_S|^2
\label{eq:TDIAGS}
\end{eqnarray}
where the upper sign applies to the vector equations
and the lower sign to the axivector equations.   The nonzero off diagonal 
$T$'s  for the vector case are
\begin{eqnarray}
T_{ab}&=T_{ba}&= \frac{i}{2}(\sigma_V^{\ast}\sigma_S-\sigma_S^{\ast}
\sigma_V)M - (\sigma_V^{\ast}\sigma_S+\sigma_S^{\ast}\sigma_V)q_3 \nonumber \\
T_{ac}&=T_{ca}&=\mbox{$\left| {\bf q} \right|$}(\sigma_V^{\ast}\sigma_S+
\sigma_S^{\ast}\sigma_V) \nonumber \\
T_{ad}&=-T_{da}&=M\mbox{$\left| {\bf q} \right|$} |\sigma_V|^2 \nonumber \\   
T_{bc}&=T_{cb}&=2 q_3 \mbox{$\left| {\bf q} \right|$} |\sigma_V|^2 \nonumber \\   
T_{bd}&=T_{db}&=i\mbox{$\left| {\bf q} \right|$} (\sigma_V^{\ast}\sigma_S
-\sigma_S^{\ast}\sigma_V) \nonumber \\   
T_{cd}&=T_{dc}&= \frac{1}{2}(\sigma_V^{\ast}\sigma_S+\sigma_S^{\ast}      
\sigma_V)M + i(\sigma_V^{\ast}\sigma_S-\sigma_S^{\ast}\sigma_V)q_3 \nonumber \\
T_{ce}&  &= M\mbox{$\left| {\bf q} \right|$} |\sigma_V|^2 \nonumber \\  
T_{cf}&  &= i\mbox{$\left| {\bf q} \right|$} (\sigma_V^{\ast}\sigma_S 
-\sigma_S^{\ast}\sigma_V) \nonumber \\  
T_{cg}&=T_{gc}&= -\frac{1}{2}(\sigma_V^{\ast}\sigma_S+\sigma_S^{\ast}      
\sigma_V)M - i(\sigma_V^{\ast}\sigma_S-\sigma_S^{\ast}\sigma_V)q_3 \nonumber \\
T_{ch}&=T_{hc}-2\mbox{$\left| {\bf q} \right|$}^2 |\sigma_V|^2&= 
-(\frac{1}{4}M^2 + q_3^{\,2} - \mbox{$\left| {\bf q} \right|$}^2) 
|\sigma_V|^2 - |\sigma_S|^2 \nonumber \\
T_{de}& &= \mbox{$\left| {\bf q} \right|$}(\sigma_V^{\ast}\sigma_S
+\sigma_S^{\ast}\sigma_V) \nonumber \\
T_{df}& &= 2\mbox{$\left| {\bf q} \right|$}q_3 |\sigma_V|^2 \nonumber \\
T_{dg}&=T_{gd}+2\mbox{$\left| {\bf q} \right|$}^2 |\sigma_V|^2&=     
-(\frac{1}{4}M^2 + q_3^{\,2} - \mbox{$\left| {\bf q} \right|$}^2) 
|\sigma_V|^2 - |\sigma_S|^2 \nonumber \\
T_{dh}&=T_{hd}&= -\frac{1}{2}(\sigma_V^{\ast}\sigma_S+\sigma_S^{\ast}
\sigma_V)M - i(\sigma_V^{\ast}\sigma_S-\sigma_S^{\ast}\sigma_V)q_3 \nonumber \\
T_{ef}&=T_{fe}&=\frac{i}{2}(\sigma_V^{\ast}\sigma_S-\sigma_S^{\ast}
\sigma_V)M - (\sigma_V^{\ast}\sigma_S+\sigma_S^{\ast}\sigma_V)q_3 \nonumber \\
T_{eg}&=T_{ge}&=-\mbox{$\left| {\bf q} \right|$}(\sigma_V^{\ast}\sigma_S
+\sigma_S^{\ast}\sigma_V) \nonumber \\
T_{eh}&=T_{he}&= -M\mbox{$\left| {\bf q} \right|$} |\sigma_V|^2 \nonumber \\
T_{fg}&=T_{gf}&= -2\mbox{$\left| {\bf q} \right|$}q_3 |\sigma_V|^2 \nonumber \\   
T_{fh}&=T_{hf}&=-i\mbox{$\left| {\bf q} \right|$}(\sigma_V^{\ast}\sigma_S
-\sigma_S^{\ast}\sigma_V) \nonumber \\
T_{ga}& &= -M\mbox{$\left| {\bf q} \right|$} |\sigma_V|^2 \nonumber \\  
T_{gb}& &=-i\mbox{$\left| {\bf q} \right|$}(\sigma_V^{\ast}\sigma_S
-\sigma_S^{\ast}\sigma_V) \nonumber \\
T_{gh}&=T_{hg}&= \frac{1}{2}(\sigma_V^{\ast}\sigma_S+\sigma_S^{\ast}
\sigma_V)M + i(\sigma_V^{\ast}\sigma_S-\sigma_S^{\ast}\sigma_V)q_3 \nonumber \\
T_{ha}& &=-\mbox{$\left| {\bf q} \right|$}(\sigma_V^{\ast}\sigma_S
+\sigma_S^{\ast}\sigma_V) \nonumber \\
T_{hb}& &= -2\mbox{$\left| {\bf q} \right|$}q_3 |\sigma_V|^2 
\end{eqnarray}
and for the axivector case the nonzero elements are
\begin{eqnarray}
T_{ab}&=-T_{ba}&= \frac{1}{2}(\sigma_V^{\ast}\sigma_S+\sigma_S^{\ast}
\sigma_V)M + i(\sigma_V^{\ast}\sigma_S-\sigma_S^{\ast}\sigma_V)q_3 \nonumber \\
T_{ac}&=-T_{ca}&=-i\mbox{$\left| {\bf q} \right|$}(\sigma_V^{\ast}\sigma_S-
\sigma_S^{\ast}\sigma_V) \nonumber \\
T_{ad}&=-T_{da}&=M\mbox{$\left| {\bf q} \right|$} |\sigma_V|^2 \nonumber \\   
T_{bc}&=T_{cb}&=2 q_3 \mbox{$\left| {\bf q} \right|$} |\sigma_V|^2 \nonumber \\   
T_{bd}&=T_{db}&=-\mbox{$\left| {\bf q} \right|$} (\sigma_V^{\ast}\sigma_S
+\sigma_S^{\ast}\sigma_V) \nonumber \\   
T_{cd}&=-T_{dc}&= \frac{i}{2}(\sigma_V^{\ast}\sigma_S-\sigma_S^{\ast}      
\sigma_V)M - (\sigma_V^{\ast}\sigma_S+\sigma_S^{\ast}\sigma_V)q_3 \nonumber \\
T_{ce}& &= -M\mbox{$\left| {\bf q} \right|$} |\sigma_V|^2 \nonumber \\  
T_{cf}& &= -\mbox{$\left| {\bf q} \right|$} (\sigma_V^{\ast}\sigma_S 
+\sigma_S^{\ast}\sigma_V) \nonumber \\  
T_{cg}&=-T_{gc}&= -\frac{i}{2}(\sigma_V^{\ast}\sigma_S-\sigma_S^{\ast}      
\sigma_V)M + (\sigma_V^{\ast}\sigma_S+\sigma_S^{\ast}\sigma_V)q_3 \nonumber \\
T_{ch}&=T_{hc}-2\mbox{$\left| {\bf q} \right|$}^2 |\sigma_V|^2& = 
-(\frac{1}{4}M^2 + q_3^{\,2} + \mbox{$\left| {\bf q} \right|$}^2) 
|\sigma_V|^2 + |\sigma_S|^2 \nonumber \\
T_{de}&  &= i\mbox{$\left| {\bf q} \right|$}(\sigma_V^{\ast}\sigma_S
-\sigma_S^{\ast}\sigma_V) \nonumber \\
T_{df}&  &= 2\mbox{$\left| {\bf q} \right|$}q_3 |\sigma_V|^2 \nonumber \\
T_{dg}&=T_{gd}+2\mbox{$\left| {\bf q} \right|$}^2 |\sigma_V|^2&=     
-(\frac{1}{4}M^2 + q_3^{\,2} - \mbox{$\left| {\bf q} \right|$}^2) 
|\sigma_V|^2 + |\sigma_S|^2 \nonumber \\
T_{dh}&=-T_{hd}&= \frac{i}{2}(\sigma_V^{\ast}\sigma_S-\sigma_S^{\ast}
\sigma_V)M - (\sigma_V^{\ast}\sigma_S+\sigma_S^{\ast}\sigma_V)q_3 \nonumber \\
T_{ef}&=-T_{fe}&=\frac{1}{2}(\sigma_V^{\ast}\sigma_S+\sigma_S^{\ast}
\sigma_V)M + i(\sigma_V^{\ast}\sigma_S-\sigma_S^{\ast}\sigma_V)q_3 \nonumber \\
T_{eg}&=-T_{ge}&=i\mbox{$\left| {\bf q} \right|$}(\sigma_V^{\ast}\sigma_S
-\sigma_S^{\ast}\sigma_V) \nonumber \\
T_{eh}&=T_{he}&=M\mbox{$\left| {\bf q} \right|$} |\sigma_V|^2 \nonumber \\
T_{fg}&=T_{gf}&=-2\mbox{$\left| {\bf q} \right|$}q_3 |\sigma_V|^2 \nonumber \\   
T_{fh}&=-T_{hf}&=-\mbox{$\left| {\bf q} \right|$}(\sigma_V^{\ast}\sigma_S
+\sigma_S^{\ast}\sigma_V) \nonumber \\
T_{ga}& &=-M\mbox{$\left| {\bf q} \right|$} |\sigma_V|^2 \nonumber \\  
T_{gb}& &=-\mbox{$\left| {\bf q} \right|$}(\sigma_V^{\ast}\sigma_S
+\sigma_S^{\ast}\sigma_V) \nonumber \\
T_{gh}&=-T_{hg}&=-\frac{i}{2}(\sigma_V^{\ast}\sigma_S-\sigma_S^{\ast}
\sigma_V)M + (\sigma_V^{\ast}\sigma_S+\sigma_S^{\ast}\sigma_V)q_3 \nonumber \\
T_{ha}& &=-i\mbox{$\left| {\bf q} \right|$}(\sigma_V^{\ast}\sigma_S
-\sigma_S^{\ast}\sigma_V)i \nonumber \\
T_{hb}& &=-2\mbox{$\left| {\bf q} \right|$}q_3 |\sigma_V|^2. 
\end{eqnarray}

%------------------------------------------------------------------
%   ACKNOWLEDGMENTS
\section*{Acknowledgments}
We are grateful to the National Centre for Theoretical Physics at 
the Australian National University where part of this work was undertaken.

%--------------------------------------------------------------------------
%   REFERENCES

%-----------------------------------------------------------------------
%   TABLES

\pagebreak

%------------------------------------------
%+++
\begin{table} 
\caption{Summary of all possible scalar and vector states 
in the nonrelativistic limit according to the generalised Pauli exclusion 
principle.}\label{tab1}
\begin{center}
\begin{tabular}{|c|c|c|c|c|}\hline
\rule{0mm}{5mm} {$J$} & {$L$} & {$S$} & {$J^{AC}$} & State  \\ \hline
\rule{0mm}{5mm} $0$ & $0$ & $0$ & $0^{-+}$ & Axiscalar  \\ 
\rule{0mm}{5mm}     & $1$ & $1$ & $0^{++}$ & Scalar     \\ \hline
\rule{0mm}{5mm} $1$ & $0$ & $1$ & $1^{--}$ & Vector     \\ 
\rule{0mm}{5mm}     & $1$ & $0$ & $1^{+-}$ & Axivector  \\ 
\rule{0mm}{5mm}     &     & $1$ & $1^{++}$ & Axivector  \\
\rule{0mm}{5mm}     & $2$ & $1$ & $1^{--}$ & Vector     \\ \hline
\end{tabular}
\end{center}
\end{table}
%
%+++
\begin{table}
\caption{Bound state masses $M$ ($\pm 0.001$) for fermion 
masses from 0 to 5.}\label{tab3} 
\begin{center}
\begin{tabular}{|c|c|c|c|c|}\hline
\rule{0mm}{5mm}
        & {Vector ${\cal C}=+1$} & {Vector ${\cal C}=-1$} & 
       {Axivector ${\cal C}=+1$} & {Axivector ${\cal C}=-1$} \\ 
\rule{0mm}{5mm} $m$ & $J^{AC} = 1^{-+}$ & $1^{--}$ & $1^{++}$ & $1^{+-}$ \\ 
       \hline
\rule{0mm}{5mm}   0   & 0.104 & 0.074 & 0.119 & 0.121 \\ \hline
\rule{0mm}{5mm} 0.001 & 0.112 & 0.081 & 0.128 & 0.129 \\ \hline
\rule{0mm}{5mm} 0.004 & 0.134 & 0.099 & 0.152 & 0.155 \\ \hline
\rule{0mm}{5mm} 0.009 & 0.160 & 0.122 & 0.179 & 0.182 \\ \hline
\rule{0mm}{5mm} 0.016 & 0.194 & 0.151 & 0.217 & 0.223 \\ \hline
\rule{0mm}{5mm} 0.025 & 0.239 & 0.184 & 0.260 & 0.282 \\ \hline
\rule{0mm}{5mm} 0.036 & 0.281 & 0.219 & 0.295 & \\ \hline
\rule{0mm}{5mm} 0.049 & 0.326 & 0.258 & 0.331 & \\ \hline
\rule{0mm}{5mm} 0.064 & 0.372 & 0.299 & 0.369 & \\ \hline
\rule{0mm}{5mm} 0.081 & 0.418 & 0.343 & 0.410 & \\ \hline
\rule{0mm}{5mm}  0.1  & 0.465 & 0.389 & 0.454 & \\ \hline
\rule{0mm}{5mm}  0.5  & 1.341 & 1.252 & 1.316 & \\ \hline
\rule{0mm}{5mm}   1   & 2.325 & 2.236 & 2.294 & \\ \hline
\rule{0mm}{5mm}   2   & 4.344 & 4.236 & 4.301 & \\ \hline
\rule{0mm}{5mm}   3   & 6.357 & 6.235 & 6.305 & \\ \hline
\rule{0mm}{5mm}   4   & 8.385 & 8.253 & 8.326 & \\ \hline
\rule{0mm}{5mm}   5   & 10.370 & 10.232 & 10.306 & \\ \hline
\end{tabular}
\end{center}
\end{table}
%
%+++
\begin{table} 
\caption{Comparison between the calculated spectrum of 4-component QED$_3$ 
positronium states in the chiral limit $m=0$ and low lying scalar 
and vector meson masses from Ref.~\protect\cite{PRG96}. The $0^{-+}$ 
QED$_3$ state is a Goldstone boson generated by the breaking of a global 
$U(2)$ symmetry. }
\label{tabmes}
\begin{center}
\begin{tabular}{|c|c|c|}\hline
\rule{0mm}{5mm}  $J^{AC}$ or $J^{PC}$    & Positronium mass $M$ & 
        Meson  \\ \hline
\rule{0mm}{5mm} $0^{-+}$ & 0     & $\pi^0$(135)     \\ 
\rule{0mm}{5mm} $0^{++}$ & 0.077 & $f_0$(400-1200)\\ 
\rule{0mm}{5mm} $0^{--}$ & 0.111 &                  \\ 
\rule{0mm}{5mm} $0^{--}$ & 0.123 &                  \\ 
\rule{0mm}{5mm} $1^{--}$ & 0.074 & $\rho$(770), $\omega$(782) \\ 
\rule{0mm}{5mm} $1^{-+}$ & 0.104 &                  \\ 
\rule{0mm}{5mm} $1^{++}$ & 0.119 & $a_1$(1260), $f_1$(1285) \\ 
\rule{0mm}{5mm} $1^{+-}$ & 0.121 & $h_1$(1170), $b_1$(1235) \\  \hline
\end{tabular}
\end{center}
\end{table}
%
%+++
\begin{table}
\caption{Charge parities of the Dirac coefficients in   
the vector, pseudovector, axivector and axipseudovector vertex functions.
The coefficient functions $f_n$ are related to the functions 
$a,b,c,d,e,f,g$ and $h$ by Eq.~(\ref{eq:RESV}) and 
(\ref{eq:RESAV}).}\label{tab5}
\begin{center}
\begin{tabular}{|c|c|c|c|}\hline
\rule{0mm}{5mm}  & {${\cal C}$} & {Even functions of $q.P$} & 
{Odd functions of $q.P$} \\ \hline
\rule{0mm}{5mm} Vector          & $+1$ & $f_2,f_6$ & $f_1,f_3,f_4,f_5,f_7,f_8$ \\ 
\rule{0mm}{5mm}                 & $-1$ & $f_1,f_3,f_4,f_5,f_7,f_8$ & $f_2,f_6$ \\ \hline
\rule{0mm}{5mm} Pseudovector    & $+1$ & $f_2,f_6$ & $f_1,f_3,f_4,f_5,f_7,f_8$ \\ 
\rule{0mm}{5mm}                 & $-1$ & $f_1,f_3,f_4,f_5,f_7,f_8$ & $f_2,f_6$ \\ \hline
\rule{0mm}{5mm} Axivector       & $+1$ & $f_1,f_2,f_4,f_7$ & $f_3,f_5,f_6,f_8$ \\ 
\rule{0mm}{5mm}                 & $-1$ & $f_3,f_5,f_6,f_8$ & $f_1,f_2,f_4,f_7$ \\ \hline
\rule{0mm}{5mm} Axipseudovector & $+1$ & $f_3,f_5,f_6,f_8$ & $f_1,f_2,f_4,f_7$ \\ 
\rule{0mm}{5mm}                 & $-1$ & $f_1,f_2,f_4,f_7$ & $f_3,f_5,f_6,f_8$ \\ \hline
\end{tabular}
\end{center}
\end{table}                           
%
%--------------------------------------------------------------------------
%   FIGURES
\section*{Figure captions.}
\begin{description}
  \vspace*{5mm}
  \item[Figure 1:] Bound state masses $M$ against fermion mass 
  $m=$ 0 to 0.1.  
Full BSE solutions from Table~\ref{tab1}: $J^{AC} = 1^{-+}$ ($\Diamond$), 
$1^{--}$ ($\Box$), $1^{+-}$ ($+$)
and $1^{++}$ ($\times$) states.

  \vspace*{5mm}
  \item[Figure 2:] Bound state masses 
$M-2m$ for fermion mass $m=$ 0 to 1. 
Full BSE solutions from Table~\ref{tab1}: $1^{-+}$ ($\Diamond$ solid), 
$1^{--}$ ($\Box$ solid), $1^{+-}$ ($+$ solid) and  $1^{++}$ ($\times$ solid).
Nonrelativistic predictions of Eq.~(\ref{eq:THY1}) using Eq.~(\ref{eq:LPRIME}): 
(Dashed curves from bottom to top) 
$\lambda'=$ 1.7969 ($\ell = 0$), 2.5666 ($\ell = 1$), 2.9316 ($\ell = 0$) 
and 3.1148 ($\ell = 2$) respectively.
  
\end{description}

%--------------------------------------------------------------------------


\begin{references}
%
\bibitem{Bu92} C.\ J.\ Burden, Nucl.\ Phys.
                        {\bf B387} 419 (1992).
%
\bibitem{AB96} T.\ W.\ Allen, and.\  C.\ J.\ Burden,  Phys.\ Rev.
                        {\bf D53} 5842 (1996).
%
\bibitem{DS79} R.\ Delbourgo, and.\ M.\ Scadron, J.\ Phys.
                        {\bf G5} 1621 (1979).
%
\bibitem{M94} P.\ Maris, Phys.\ Rev.
                        {\bf D50} 4189 (1994).
%
\bibitem{M95} P.\ Maris, Phys.\ Rev and references therein.
                        {\bf D52} 6087 (1995).
%
\bibitem{SC92} S.\ J.\ Stainsby, and.\  R.\ T.\ Cahill,  
      Int.\ J.\ Mod.\ Phys. {\bf A7} 7541 (1992); S.\ J.\ Stainsby, and.\  
       R.\ T.\ Cahill,  Mod.\ Phys.\ Lett. {\bf A9} 3551 (1994).  
%
\bibitem{Sev} C.\ J.\ Burden, and.\  C.\ D.\ Roberts, 
              Phys.\ Rev. {\bf D44} 540 (1991); 
              M.\ R.\ Pennington and   D.\ Walsh,  
              Phys.\ Lett. {\bf B253} 246 (1991);  
              D.\ C.\ Curtis, M.\ R.\ Pennington and.\ D.\ Walsh,  
              Phys.\ Lett. {\bf B295} 313 (1992).  
%
\bibitem{BRS96} A.\ Bender, C.\ D.\ Roberts and.\ L.\ von Smekal,   
                Phys.\ Lett. {\bf B380} 7 (1996).
%
\bibitem{N94} M.\ Neubert, Phys. Rep. {\bf 245} 259 (1994).  
%
\bibitem{Ko96} V.\ G.\ Koures, J.\ Comp.\ Phys. {\bf 128} 1 (1996). 
%
\bibitem{KN86} Y.\ S.\ Kim and M.\ E.\ Noz, {\it Theory and Applications of 
the Poincare Group}, Dordrecht Holland, (1986).  
%
\bibitem{FS82} D.\ Flamm and F.\ Sch\"{o}berl, {\it Introduction to the
Quark Model of Elementary Particles}, Gordon and Breach, (1982).  
%
\bibitem{THY95} A.\ Tam, C.\ J.\ Hamer and C.\ M.\ Yung, J. Phys. 
               {\bf G21} 1463 (1995).  
%
\bibitem{PRG96} R.\ M.\ Barnett et al.  (Particle Physics Group) 
    Phys.\ Rev. {\bf D54} 1 (1996).
%
\bibitem{BB87} C.\ J.\  Burden and A.\ N.\ Burkitt,    
                Europhys.\ Lett. {\bf 3} 545 (1987).
%
\bibitem{DKK90} E.\ Dagotto et al., Nucl. Phys. {\bf B334} 279 (1990). 
%
\bibitem{B92} C.\ Burden, How can QED$_3$ help us understand QCD$_4$, 
              in {\it QCD Vacuum Structure}, edited by H. M. Fried and 
              B. M\"uller, World Scientific, Singapore, 1993.   
%
\bibitem{LS69} C.\ H.\  Llewelyn-Smith, Ann. Phys. {\bf 53} 521 (1969).  
%
\end{references}
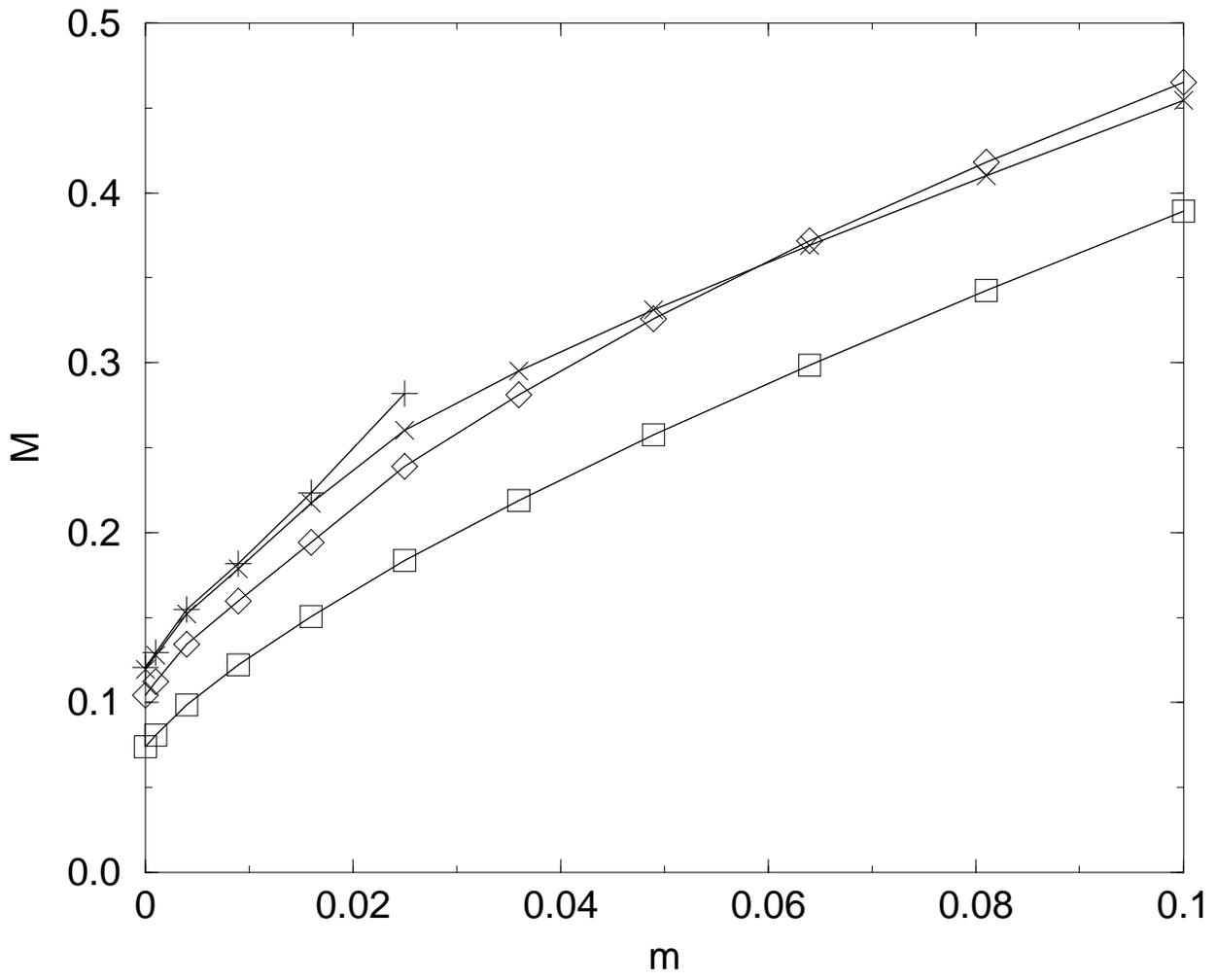
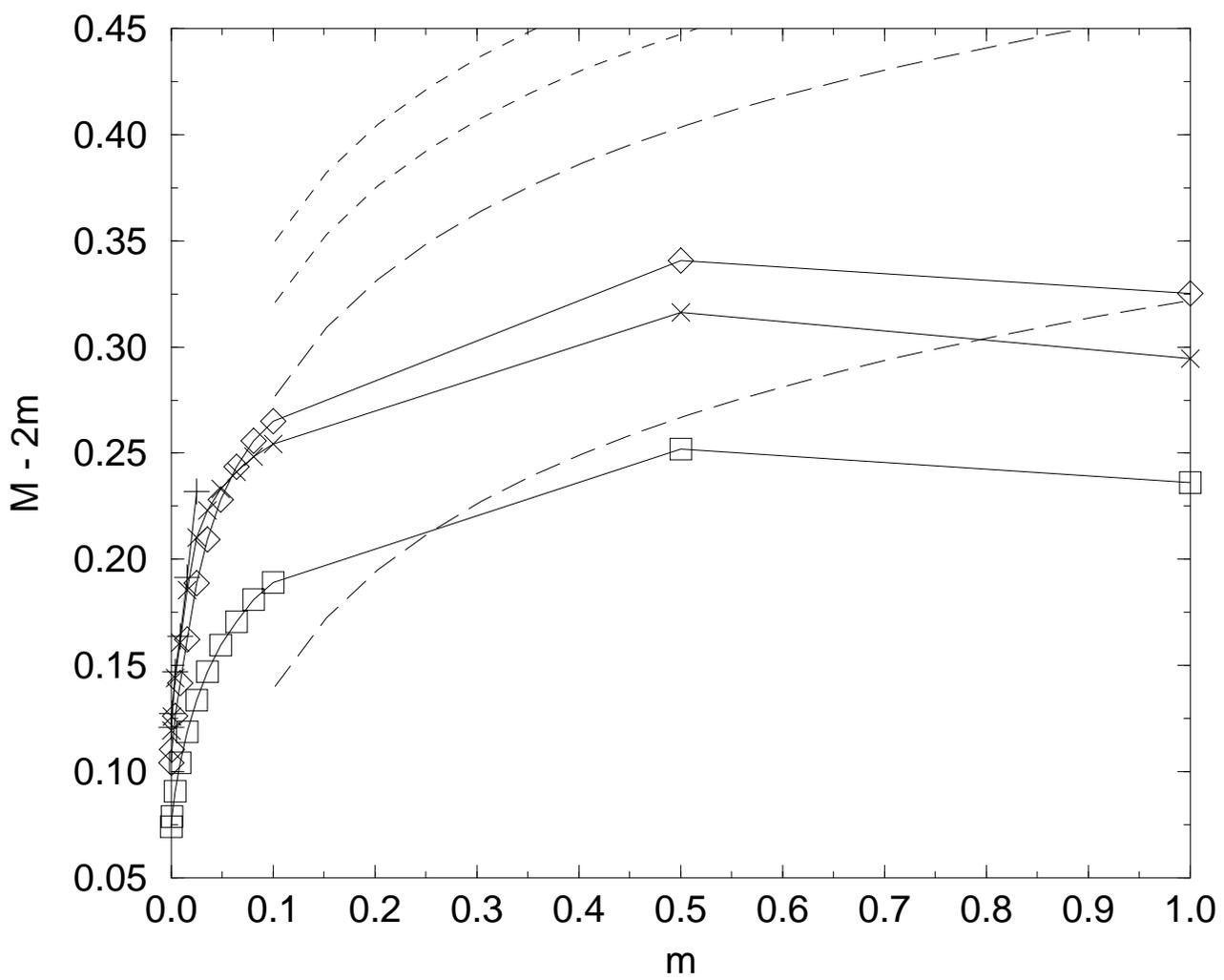
\end{document}